\documentclass[12pt]{article}
\usepackage[T1]{fontenc}
\usepackage{amsmath}
\DeclareMathAlphabet{\mathcal}{OMS}{cmsy}{m}{n}
\usepackage{amsthm}
\usepackage{amssymb}
\usepackage{txfonts}
\usepackage{mathrsfs}
\usepackage{graphicx}
\usepackage{cite}
\usepackage{bm}
\usepackage{xcolor}
\usepackage{tikz}
\usetikzlibrary{decorations.pathreplacing}
\tikzset{every picture/.style={remember picture}}
\usepackage{booktabs}
\usepackage{enumitem}
\usepackage{xurl}
\usepackage{hyperref}
\usepackage{float}
\hypersetup{hidelinks,pdftitle={Constructions of optimal Ferrers diagram rank-metric codes for quasi-block monotone Ferrers diagrams}}
\setlist[itemize]{itemsep=2pt,parsep=0pt,topsep=4pt,partopsep=0pt}
\setlist[enumerate]{label=\textup{(\arabic*)},itemsep=0.15em,topsep=0.3em}
\newtheorem{definition}{Definition}[section]
\newtheorem{proposition}[definition]{Proposition}
\newtheorem{theorem}[definition]{Theorem}
\newtheorem{lemma}[definition]{Lemma}
\newtheorem{corollary}[definition]{Corollary}
\newtheorem{notation}[definition]{Notation}
\newtheorem{remark}[definition]{Remark}
\newtheorem{example}[definition]{Example}
\DeclareMathOperator{\supp}{supp}
\newcommand{\F}{\mathbb F}
\newcommand{\C}{\mathcal C}

\newcommand{\rank}{\operatorname{rank}}

\newcommand{\nuMin}{\nu_{\min}}
\numberwithin{equation}{section}
\allowdisplaybreaks[2]
\begin{document}
\begin{center}
{\Large\bfseries New constructions of optimal Ferrers diagram rank-metric codes for quasi-block monotone Ferrers diagrams\par}
\end{center}
\centerline{ {\sc Yong Jin$^1$\qquad Yuhao Ning$^2$ \qquad Dengming Xu$^3$ \qquad Chengming Duan$^4$   }}
\begin{abstract}
Constant dimension subspace codes (CDCs) have become an important object in coding theory due to their applications in random network coding. The paper is devoted to constructing optimal Ferrers diagram rank-metric codes for block Ferrers diagrams, which play a crucial role in the multilevel construction of CDCs. 
Let \(\mathbb{F}_q\) be the finite field with \(q\) elements of characteristic \(p\), \(P\) be a power of \(p\), and \(e\) be a positive integer. Let \(L=\mathbb F_{q^{eP}}\) and \(\sigma\) be the Frobenius \(\mathbb F_q\)-automorphism of \(L\), set \(N=\sigma^e-\mathrm{id}\). 
Based on the multiplicative property of the kernel filtration of the nilpotent map \(N\), we construct optimal Ferrers diagram rank-metric codes for block monotone diagrams. Specifically, considering \(n \times l\) Ferrers diagrams, we prove that optimal codes always exist for block monotone diagrams when \(n = eP\). Furthermore, by employing puncturing on columns, we prove that optimal codes also always exist for block strictly monotone diagrams, removing the restriction \(n = eP\). Consequently, optimal codes always exist for block upper triangular diagrams, specifically for maximum
sum-rank distance (MSRD)-constructible block diagram pairs, which generalizes the corresponding results in \cite{neri2024,calderini2026}, in which the rank distance or the characteristic is restricted. Furthermore, by employing both puncturing and shortening on columns, we prove that optimal codes always exist for quasi-block strictly monotone diagrams, thereby generalizing the corresponding results in \cite{pratihar2023, zhang2019}. The optimal codes constructed in this paper will potentially be beneficial to the further construction of CDCs.
  
\end{abstract}

\renewcommand{\thefootnote}{\alph{footnote}}
\setcounter{footnote}{-1} \footnote{$^1$ Corresponding author, College of Science,
	Civil Aviation University of China, 300300 Tianjin,
	PR  China. E-mail: kingmeng@126.com.
}
\setcounter{footnote}{-1} \footnote{$^2 $ College of Science,
	Civil Aviation University of China, 300300 Tianjin,
	PR  China. E-mail: ningyuh00@163.com
}
\setcounter{footnote}{-1} \footnote{$^3 $ Sino-European Institute of Aviation Engineering,
Civil Aviation University of China, 300300 Tianjin,
PR China. E-mail:
xudeng17@163.com 
}
\setcounter{footnote}{-1} \footnote{$^4 $ College of Science,
Civil Aviation University of China, 300300 Tianjin,
PR  China. E-mail: 19160306346@163.com 
}
\par\smallskip
\section{Introduction}\label{sec:intro}
In the study of random network coding, K{\"o}tter and Kschischang described the network coding model in terms of subspaces and introduced the subspace distance, establishing a standard theoretical framework for subspace codes \cite{koetter2008}. Constant dimension subspace codes (CDCs) form an important class of subspace codes for correcting errors and erasures. Lifting maximum rank distance (MRD) codes yields constant dimension subspace codes with good asymptotic performance \cite{silva2008,EtzionSilberstein2013}. Etzion and Silberstein \cite{etzion2009} generalized this approach by introducing Ferrers diagram rank-metric codes and proposing the multilevel construction of CDCs. They established an upper bound on the dimension of these codes and conjectured that the bound is attainable for every choice of parameters. Ferrers diagram rank-metric codes attaining the dimension bound are called optimal. Subsequent improvements of the multilevel construction have produced CDCs with larger cardinalities \cite{hong2025,liu2020parallel,liu2023double,silberstein2015,yu2022bilateral}. Cosets of Ferrers diagram rank-metric codes have also been used to construct CDCs \cite{XuSong2025}. Optimal Ferrers diagram rank-metric codes play an important role in these constructions.

Optimal Ferrers diagram rank-metric codes have been constructed for various Ferrers diagrams and minimum distances\cite{liu2019,liu2023,etzion2016,antrobus2019,neri2024,pratihar2023,calderini2026,liu2019several,zhang2019,fu2026}. The main constructions in the literature are based on subcodes of MRD codes and combinatorial methods. Prajapati \cite{Prajapati2026} recently gave a
counterexample, showing that optimal codes do not always exist for every choice of parameters. Nevertheless, the role of Ferrers diagram rank-metric codes in the multilevel construction makes it important to determine for which Ferrers diagrams and minimum distances there exist optimal codes.

Block Ferrers diagram rank-metric codes have also been studied. Using automorphisms of rational function fields, Pratihar and Randrianarisoa constructed a class of optimal block Ferrers diagram codes under restrictions on the field size \cite{pratihar2023}. Neri and Stanojkovski proved the existence of optimal codes for strictly monotone and MDS-constructible Ferrers diagrams over every finite field. They also constructed optimal codes for block Ferrers diagrams whose block sizes depend on the characteristic $p$ \cite{neri2024}.

Calderini, Messia and Neri developed a diagonal construction based on maximum sum-rank distance (MSRD) codes\cite{calderini2026}. For Ferrers diagrams consisting of square blocks, they defined MSRD-constructible block diagram pairs and reduced their optimality over every finite field to that of block upper triangular diagrams \cite[Theorem~5.3]{calderini2026}. They constructed optimal Ferrers diagram rank-metric codes for block upper triangular diagrams with $s$ diagonal blocks of size $e$ and the following minimum distances:
\begin{equation*}\label{eq:oldrange}
 d\in\{1,\ldots,e+1\}\cup\{es-1,es\}
\end{equation*}

We continue the study of the existence of optimal Ferrers diagram rank-metric codes for block Ferrers diagrams. We introduce the notions of quasi-block monotone and quasi-block strictly monotone Ferrers diagrams (Definition~\ref{quasiblock}), generalizing the existing notions of block Ferrers diagrams. Let $\mathbb{F}_q$ be the finite field with $q$ elements of characteristic $p$, $P$ be a power of $p$, and $e$ be an arbitrary positive integer. Set $E=\F_{q^e}$ and $L=\F_{q^{eP}}$. Let \(\sigma\) be the Frobenius \(\mathbb F_q\)-automorphism of \(L\), set \(N=\sigma^e-\mathrm{id}\). We use $\sigma$ and $N$ to obtain an $L$-basis of $L[\sigma]$. Based on the multiplicative property of the kernel filtration of the nilpotent map \(N\), we construct optimal rank-metric codes for block monotone diagrams. Specifically, considering \(n \times l\) Ferrers diagrams, we prove that optimal codes always exist for block monotone diagrams when \(n = eP\) (Theorem\ref{mainthm1}). We then extend the result to block strictly monotone diagrams without the restriction $n=eP$ and show the existence of optimal codes over any finite field for every minimum rank distance (Theorem\ref{mainthm2}). As a special case, the Etzion--Silberstein conjecture holds for block upper triangular diagrams. This extends \cite[Theorem~5.9]{calderini2026} to every admissible
minimum rank distance. Together with \cite[Theorem~5.3]{calderini2026},
it implies that for all MSRD-constructible block diagram pairs,  there exist optimal codes over any finite field. As another application, for even $t$ and $q\in\{2,3\}$, we increase the dimension of a family of Ferrers diagram rank-metric codes used in the multilevel construction of Xu and Li \cite{xu2025} from $3t$ to $t^2+2t$. For block monotone diagrams with $n \neq eP$ rows, we prove the existence of optimal Ferrers diagram rank-metric codes under additional conditions.

By employing both puncturing and shortening on columns, we prove that optimal codes always exist for quasi-block strictly monotone diagrams(Theorem\ref{thm-ni}). This result removes the field-size restrictions in the construction of Pratihar and Randrianarisoa based on automorphisms of rational function fields, allowing optimal codes for these diagrams to be constructed over smaller fields. By our result we  weaken the sufficient conditions in Zhang and Ge \cite[Theorem~3.2]{zhang2019}. Moreover, we show the existence of optimal codes for certain quasi-block monotone diagrams obtained by deleting arbitrary columns from block monotone diagrams covered by Theorem\ref{mainthm1} or Proposition
\ref{prop-mon-block}.

We give several examples of optimal codes for diagrams not covered by previous constructions. We also pose open problems on the existence of optimal codes for block Ferrers diagrams that remain unresolved by both our methods and earlier constructions.

This paper is organized as follows. 
In Section~\ref{sec:prelim}, we introduce the notation and give some necessary definitions and results used in our main constructions. 
In Section~\ref{sec:construction}, we give our main constructions of optimal codes and give their applications. Section~\ref{sec:conclusion} concludes the paper.

\section{Preliminaries}\label{sec:prelim}
For a positive integer \(n\), write
\[
[n]=\{1,\ldots,n\},
\qquad
[n]\times[l]
=\{(i,j):i\in[n],j\in[l]\}
\]
and let \([0]=\varnothing\).

\subsection{Ferrers diagram rank-metric codes}

\begin{definition}\label{Ferrersdiag}
Let $n$ and $l$ be positive integers. An $n\times l$ \textbf{Ferrers diagram} $\mathcal{D}$ is an $n\times l$ array of dots and empty entries satisfying the following conditions:
\begin{itemize}
  \item each row contains at most as many dots as the preceding row;
  \item all dots are aligned to the right.

\end{itemize}
\end{definition}

An $n\times l$ Ferrers diagram $\mathcal{D}$ can also be identified with a subset of $[n]\times[l]$, where $(i,j)\in \mathcal{D}$ if and only if there is a dot in position $(i,j)$. This subset is called the support of $\mathcal{D}$ and is denoted by $\supp (\mathcal{D})$.

Let $\mathcal{D}$ be an $n\times l$ Ferrers diagram. For $1\le i\le l$, let $c_i$ be the number of dots in the $i$th column of $\mathcal{D}$. Then $\mathcal{D}$ can be written as
\[
\mathcal{D}=[c_1,c_2,\ldots,c_l].
\]
The transposed Ferrers diagram of $\mathcal{D}$ is defined by
\[
\mathcal{D}^{t}=[\rho_n,\rho_{n-1},\ldots,\rho_1],
\]
where $\rho_i$ is the number of dots in the $i$th row of $\mathcal{D}$, $1\le i\le n$. Write $|\mathcal{D}|=\sum_{j=1}^{l}c_j$. An $n\times l$ Ferrers diagram is called full if it contains $nl$ dots.

\begin{example}
The $5\times 4$ Ferrers diagram $\mathcal{D}=[1,2,4,5]$ and its transpose $\mathcal{D}^{t}=[1,2,2,3,4]$ are as follows:
\[
\renewcommand{\arraystretch}{0.8}
\mathcal{D}=\quad
\begin{array}{c@{\quad}c@{\quad}c@{\quad}c}
\bullet & \bullet & \bullet & \bullet \\
        & \bullet & \bullet & \bullet \\
        &         & \bullet & \bullet \\
        &         & \bullet & \bullet \\
        &         &         & \bullet
\end{array}
\qquad
\mathcal{D}^{t}=\quad
\begin{array}{c@{\quad}c@{\quad}c@{\quad}c@{\quad}c}
\bullet & \bullet & \bullet & \bullet & \bullet \\
        & \bullet & \bullet & \bullet & \bullet \\
        &         &         & \bullet & \bullet \\
        &         &         &         & \bullet
\end{array}.
\]
\end{example}

Let $\mathcal D$ and $\mathcal{D}'$ be $n\times l$ Ferrers diagrams. If $\supp(\mathcal{D}')\subseteq\supp(\mathcal{D})$, then $\mathcal{D}'$ is called a subdiagram of $\mathcal{D}$, denoted by $\mathcal{D}'\subseteq\mathcal{D}$.
\begin{example}
Let $\mathcal D=[2,3,5,5]$ and $\mathcal D'=[0,2,5,5]$. Their diagrams are
\[
\renewcommand{\arraystretch}{0.8}
\mathcal D=\quad
\begin{array}{c@{\quad}c@{\quad}c@{\quad}c}
\bullet & \bullet & \bullet & \bullet \\
\bullet & \bullet & \bullet & \bullet \\
        & \bullet & \bullet & \bullet \\
        &         & \bullet & \bullet \\
        &         & \bullet & \bullet
\end{array}
\qquad
\mathcal D'=\quad
\begin{array}{c@{\quad}c@{\quad}c}
\bullet & \bullet & \bullet \\
\bullet & \bullet & \bullet \\
        & \bullet & \bullet \\
        & \bullet & \bullet \\
        & \bullet & \bullet
\end{array}.
\]
Thus $\mathcal D'\subseteq\mathcal D$.
\end{example}

Let $\mathbb{F}_q$ be a finite field with $q$ elements, and let $\mathbb{F}_q^{n\times l}$ denote the space of all $n\times l$ matrices over $\mathbb{F}_q$. The rank of $A\in\mathbb{F}_q^{n\times l}$ is denoted by $\operatorname{rank}(A)$.
The rank distance on $\mathbb{F}_q^{n\times l}$ is defined by
\[
d_R(A,B)=\operatorname{rank}(A-B),
\qquad A,B\in\mathbb{F}_q^{n\times l}.
\]
An $(n\times l,M,d)_q$ rank-metric code is a subset $\mathcal{C}\subseteq\mathbb{F}_q^{n\times l}$ of cardinality $M$ with minimum rank distance
\[
d=\min_{\substack{A,B\in\mathcal{C}\\A\neq B}}d_R(A,B).
\]

For every rank-metric code, its cardinality satisfies the Singleton-type bound\cite{delsarte}
\[
M\leq q^{\max\{l,n\}(\min\{l,n\}-d+1)}.
\]
A code attaining this bound is called a \textbf{maximum rank distance} (MRD) code. As fundamental objects in rank-metric code theory, MRD codes provide a crucial starting point to construct optimal Ferrers diagram rank-metric codes.

Furthermore, if $\mathcal{C}$ is a $k$-dimensional $\mathbb{F}_q$-linear subspace of $\mathbb{F}_q^{n\times l}$, then it is called a linear rank-metric code and is denoted by $[n\times l,k,d]_q$. Linear MRD codes exist over every finite field $\mathbb{F}_q$ for every $1\le d\le \min\{n,l\}$ \cite{delsarte,gabidulin}.

Let $\mathcal{D}$ be an $n\times l$ Ferrers diagram. A matrix in $\mathbb{F}_q^{n\times l}$ is said to have shape $\mathcal{D}$ if all its entries outside $\supp(\mathcal{D})$ are zero. Let $\mathbb{F}_q^{\mathcal{D}}$ denote the set of all matrices in $\mathbb{F}_q^{n\times l}$ having shape $\mathcal{D}$. Then $\mathbb{F}_q^{\mathcal{D}}$ is an $\mathbb{F}_q$-subspace of $\mathbb{F}_q^{n\times l}$ of dimension $|\mathcal{D}|$.
An $[\mathcal{D},k,d]_q$ \textbf{Ferrers diagram rank-metric code} $\C$ is a
$k$-dimensional $\mathbb{F}_q$-linear subspace of $\mathbb{F}_q^{\mathcal{D}}$
whose minimum rank distance is defined by
\[
d=\min\{\rank(A-B):A,B\in\C,\ A\ne B\}=\min\{\rank(A):A\in\C,\ A\ne0\}.
\]

\begin{lemma}(Etzion--Silberstein bound, \cite[Theorem~1]{etzion2009})\label{singleton-bound}
Let $\mathcal{D}=[c_1,\ldots,c_l]$ and $1\le d\le l$. For
$0\le t\le d-1$, set
\[
\nu_t(\mathcal{D},d)
=\sum_{j=1}^{l-d+1+t}\max\{0,c_j-t\},
\qquad
\nu_{\min}(\mathcal{D},d)
=\min_{0\le t\le d-1}\nu_t(\mathcal{D},d).
\]
Then every $[\mathcal{D},k,d]_q$ code satisfies
\[
k\le \nu_{\min}(\mathcal{D},d).
\]
 Here, \(\nu_t(\mathcal{D},d)\) denotes the number of dots remaining after deleting the first t rows and the rightmost \(d-1-t\) columns of \(\mathcal{D}\).
\end{lemma}

\begin{example}
Let \(\mathcal{D}=[1,2,3,3,5]\) and \(d=3\). Then
\[
\nu_0(\mathcal{D},3)=6,\qquad
\nu_1(\mathcal{D},3)=5,\qquad
\nu_2(\mathcal{D},3)=5.
\]
Hence \(\nuMin(\mathcal{D},3)=5\).
\end{example}

A Ferrers diagram rank-metric code attaining the above dimension bound is called optimal. The \textbf{Etzion--Silberstein conjecture} asserts that, for every $n\times l$ Ferrers diagram $\mathcal{D}$ and every $d\in\{1,\ldots,\min\{l,n\} \}$, there exists an optimal $[\mathcal{D},\nu_{\min}(\mathcal{D},d),d]_q$ code over every finite field $\mathbb{F}_q$. Beeloo-Sauerbier Couv\'{e}e and Neri\cite{couvee2026} showed that the conjecture holds if and only if it holds for all irreducible diagram pairs $(\mathcal{D},d)$ over every finite field $\mathbb{F}_q$. Prajapati recently gave a counterexample \cite{Prajapati2026}, disproving the conjecture. Nevertheless, these codes play an essential part in the multilevel construction of constant dimension subspace codes, so it is important to determine for which parameters there exist optimal Ferrers diagram rank-metric codes. We note several existence cases that will be used in the constructions of the next section.
\begin{proposition}\cite{etzion2009}\label{fullpoints} Let $\mathcal{D}$ be an $n\times l$ Ferrers diagram with $n\ge l$, and let $1\le d\le l$. Suppose that each of the rightmost $d-1$ columns contains $n$ dots. Then there exists an optimal $[\mathcal{D},\nu_0(D,d),d]_q$ code for every $q\geq 2$.
\end{proposition}

\begin{lemma}\cite[Proposition 4.23]{neri2024}\label{pendingdot}
Let $\mathcal{D}$ be an $n\times l$ Ferrers diagram, and let $d$ be an integer with $1\le d\le \min\{n,l\}$.
\begin{enumerate}
\item[(1)]
Suppose that $\mathcal{D}'\subseteq\mathcal{D}$ and
\[
\nu_{\min}(\mathcal{D},d)=\nu_{\min}(\mathcal{D}',d).
\]
If $\mathcal{C}$ is an optimal
$[\mathcal{D}',\nu_{\min}(\mathcal{D}',d),d]_q$
code, then $\mathcal{C}$ is also an optimal
$[\mathcal{D},\nu_{\min}(\mathcal{D},d),d]_q$
code.

\item[(2)]
Suppose that $\mathcal{D}'\supseteq\mathcal{D}$ and
\[
\nu_{\min}(\mathcal{D},d)
=
\nu_{\min}(\mathcal{D}',d)-(|\mathcal{D}'|-|\mathcal{D}|).
\]
If $\mathcal{C}$ is an optimal
$[\mathcal{D}',\nu_{\min}(\mathcal{D}',d),d]_q$
code, then
$\mathcal{C}\cap\mathbb{F}_q^{\mathcal{D}}$ is an
optimal $[\mathcal{D},\nu_{\min}(\mathcal{D},d),d]_q$
code.
\end{enumerate}
\end{lemma}

By symmetry, $\nuMin(\mathcal{D},d)=\nuMin(\mathcal{D}^{t},d)$, and an $[\mathcal{D},\nuMin(\mathcal{D},d),d]_q$ code exists if and only if an optimal $[\mathcal{D}^{t},\nuMin(\mathcal{D}^{t},d),d]_q$ code exists. Therefore, without loss of generality, we assume that the first row of each \(n\times l\) Ferrers diagram has no more dots than its rightmost column throughout this paper.
\subsection{Quasi-block monotone and quasi-block strictly monotone Ferrers diagrams}

\begin{definition}\label{quasiblock}
Let \(\mathcal{D} = [c_1, c_2, \dots, c_l]\) be an \(n\times l\) Ferrers diagram. Suppose that there is a positive integer \(e \ge 1\) such that \(e \mid c_i\) for every \(i \in [l]\).
\begin{enumerate}
\item[(1)] We call \(\mathcal{D}\) \emph{\textbf{quasi \(e\)-block monotone}} if, for all \(1 \le i < j \le l\), the conditions \(c_i = c_j > 0\) and \(c_i < c_l\) imply \(j - i < e\).
\item[(2)] We call \(\mathcal{D}\) \emph{\textbf{quasi \(e\)-block strictly monotone}} if, for all \(1 \le i < j \le l\), the condition \(c_i = c_j > 0\) implies \(j - i < e\).
\end{enumerate}
\end{definition}

\begin{remark}
In \cite{calderini2026}, an \(e\)-block strictly monotone diagram consists of \(e \times e\) blocks, that is, every  group of consecutive columns of equal height contains exactly \(e\) columns. The definition above allows the number \(s\) of such columns to satisfy \(1 \le s \le e\), equivalently \(j - i < e\). Thus the \(e\)-block strictly monotone diagrams of \cite{calderini2026} are precisely the special case \(s = e\). The same observation applies to \(e\)-block monotone diagrams, so our definitions extend the existing ones.
\end{remark}

\begin{notation}
Let \(\mathcal{D}\) be a quasi \(e\)-block monotone Ferrers diagram, with \(k\) distinct positive column heights \(c_{i_1} < c_{i_2} < \dots < c_{i_k}\). Suppose that height \(c_{i_t}\) occurs in \(e_t\) consecutive columns, $t=1,2,\cdots,k$, and write \(e_k=\theta e+\gamma\), where $\theta\ge 0$ and $0\le \gamma<e$. Then \(\mathcal{D}\) can be written as follows.
\[
\mathcal{D} = [[(eh_1)^{({e_1})}, (eh_2)^{({e_2})}, \dots, (eh_{k-1})^{({e_{k-1}})},\underbrace{{(eh_k)}^{(e)},\cdots,{(eh_k)}^{(e)}}_\theta,{(eh_k)}^{(\gamma)}]]_e,
\]
where \(h_t = c_{i_t}/e\), $t=1,2,\cdots,k$, is the corresponding block height, and the superscript \(e_t\) indicates that this height occurs \(e_t\) times. If \(\mathcal{D}\) is a quasi \(e\)-block strictly monotone Ferrers diagram, then every \(e_t \le e\) and \(\mathcal{D}\) can be written as 
\[
\mathcal{D} = [[(eh_1)^{({e_1})}, (eh_2)^{({e_2})}, \dots, (eh_{k-1})^{({e_{k-1}})},{(eh_k)}^{(e_k)}]]_e. 
\]
 If every \(e_t = e\), we write \(\big[[h_1, h_2, \dots, h_k]\big]_e\).
\end{notation}

\begin{example}
The Ferrers diagram $\mathcal{D}=\big[[3^{(2)},6^{(2)},9^{(1)},12^{(3)}]\big]_3$ is quasi \(3\)-block strictly monotone:
\[
\renewcommand{\arraystretch}{0.8}
\mathcal{D}=\quad
\begin{array}{c@{\quad}c@{\quad}c@{\quad}c@{\quad}c@{\quad}c@{\quad}c@{\quad}c}
\bullet & \bullet & \bullet & \bullet & \bullet & \bullet & \bullet & \bullet \\
\bullet & \bullet & \bullet & \bullet & \bullet & \bullet & \bullet & \bullet \\
\bullet & \bullet & \bullet & \bullet & \bullet & \bullet & \bullet & \bullet \\
        &         & \bullet & \bullet & \bullet & \bullet & \bullet & \bullet \\
        &         & \bullet & \bullet & \bullet & \bullet & \bullet & \bullet \\
        &         & \bullet & \bullet & \bullet & \bullet & \bullet & \bullet \\
        &         &         &         & \bullet & \bullet & \bullet & \bullet \\
        &         &         &         & \bullet & \bullet & \bullet & \bullet \\
        &         &         &         & \bullet & \bullet & \bullet & \bullet \\
        &         &         &         &         & \bullet & \bullet & \bullet \\
        &         &         &         &         & \bullet & \bullet & \bullet \\
        &         &         &         &         & \bullet & \bullet & \bullet
\end{array}.
\]
\end{example}

\subsection{Skew group algebra}
We recall the skew group algebra $L[\sigma]$ which will be used in the constructions of Section~3. Further details can be found in \cite{neri2024}.

Let \(F=\F_{q}\), \(L=\F_{q^n}\), and define
\[
\sigma:L\longrightarrow L,\qquad x\longmapsto x^q.
\]
Then \(\operatorname{Gal}(L/F)=\langle\sigma\rangle\).
The skew group algebra \(L[\sigma]\) is defined by
\[
L[\sigma]=
\left\{
\sum_{i=0}^{n-1}a_i\sigma^i:a_i\in L
\right\}
\]
with usual addition, and multiplication determined by
\[
(a\sigma^i)(b\sigma^j)
=
a\,\sigma^i(b)\sigma^{i+j},
\qquad a,b\in L,
\]
and extended by distributivity. Here $\sigma^n=\mathrm{id}$, where $\mathrm{id}$ denotes the identity map.
The map

\[
\sum_{i=0}^{n-1}a_i\sigma^i
\longmapsto
\left(x\longmapsto\sum_{i=0}^{n-1}a_i\sigma^i(x)\right),
\]
identifies \(L[\sigma]\) with \(\operatorname{End}_F(L)\).

Let \({\mathscr{B}}=(\beta_1,\ldots,\beta_n)\) be an
ordered \(F\)-basis of \(L\). Define
\[
\phi_{\mathscr{B}}:L[\sigma]\longrightarrow F^{n\times n},
\qquad
f\longmapsto
\bigl([f(\beta_1)]_{\mathscr{B}},\ldots,
[f(\beta_n)]_{\mathscr{B}}\bigr),
\]
where \([x]_{\mathscr{B}}\) is the coordinate column vector of \(x\) with respect to \({\mathscr{B}}\).
The map \(\phi_{\mathscr{B}}\) is an \(F\)-algebra isomorphism.

For a nonzero element \(f=\sum_{i=0}^{n-1}a_i\sigma^i\), its
\(\sigma\)-degree is defined by
\[
\deg_\sigma f=\max\{i:a_i\ne0\}.
\]

\begin{lemma}\cite{GowQuinlan2009}\label{lem:degree}
For every \(0\neq f\in L[\sigma]\), we have
    \[
    \dim_F \ker f\le\deg_\sigma f,
    \qquad
    \rank\bigl(\phi_{\mathscr{B}}(f)\bigr)
    \ge n-\deg_\sigma f.
    \]

\end{lemma}

\subsection{Column puncturing and shortening of rank-metric codes}

We recall two basic operations on rank-metric codes, which are puncturing and shortening, and describe their effects on dimension and minimum rank distance of codes.

Let \(\mathcal{C} \subseteq \mathbb{F}_q^{n \times l}\) be a linear rank-metric code of dimension \(k \) and minimum rank distance \(d\).
Let \(J \subseteq [l]\) be a set of column indices with \(|J|=s\).
Write \(\overline{J} = [l] \setminus J\) for the complement of \(J\), with \(|\overline{J}| = l-s\).
For any matrix \(M \in \mathbb{F}_q^{n \times l}\), let \(M|_{\overline{J}}\) denote the \(n \times (l-s)\) matrix obtained by retaining the columns of \(M\) indexed by \(\overline{J}\).

\begin{definition}
Let $\mathcal{C}\subseteq \mathbb{F}_q^{n \times l}$. The \emph{\textbf{punctured code}} of \(\mathcal{C}\) at \(J\) is
	\[
	\mathcal{C}_{\text{punc}} = \mathcal{C}|_{\overline{J}} = \{ M|_{\overline{J}} \mid M \in \mathcal{C} \}.
	\]
Thus the columns indexed by \(J\) are deleted from every codeword of \(\mathcal{C}\). Clearly, if \(|J|=s\), then \(\mathcal{C}_{\text{punc}}\subseteq \mathbb{F}_q^{n \times {(l-s)}}\).
\end{definition}

\begin{definition}
Let $\mathcal{C}\subseteq \mathbb{F}_q^{n \times l}$. The \emph{\textbf{shortened code}} of \(\mathcal{C}\) at \(J\) is
	\[
\mathcal{C}_{ \text{short} }=\{ M|_{\overline{J}}\mid M\in\mathcal{C},\ \text{and column } j \text{ of } M \text{ is zero for every } j \in J \}.
	\]
Thus we first take the subcode of \(\mathcal{C}\) whose columns indexed by \(J\) are all zero, and then delete these zero columns to obtain $\mathcal{C}_{\text{short}}$. Clearly, if \(|J|=s\), then \(\mathcal{C}_{\text{short}}\subseteq \mathbb{F}_q^{n \times {{(l-s)}}}\).
\end{definition}

Let $\mathcal{D}$ be an $n\times l$ Ferrers diagram, and let \(J \subseteq [l]\) be a set of column indices with \(|J|=s\). Denote by $\mathcal D|_{\overline J}$ the $n\times (l-s)$ Ferrers diagram obtained by deleting the columns of $\mathcal{D}$ indexed by $J$. Let $\mathcal{C}\subseteq \mathbb{F}_q^{n \times l}$. If $M$ has shape $\mathcal{D}$ for every codeword $M\in \mathcal{C}$, then every codeword in the punctured code $\mathcal{C}_{\text{punc}}$ and the shortened code $\mathcal{C}_{\text{short}}$ of $\mathcal{C}$ at $J$ has shape $\mathcal D|_{\overline J}$.

\begin{lemma}\cite{Neri2019}\label{lem:puncture}
Let $\mathcal{C}$ be an $[n\times l,k,d]_q$ rank-metric code, and let \(J \subseteq [l]\) with \( |J| = s\). Then
	\begin{enumerate}
\item \(\dim(\mathcal{C}_{\text{punc}}) \le k\). If \(s < d\), then \(\dim(\mathcal{C}_{\text{punc}}) = k\).
\item \(d(\mathcal{C}_{\text{punc}}) \ge d - s\).
	\end{enumerate}
\end{lemma}

\begin{proof}
(1) Column deletion is a linear projection \(\Pi: \mathbb{F}_q^{n \times l} \to \mathbb{F}_q^{n \times (l-s)}\). The dimension of its image cannot exceed that of the original space, so \(\dim(\mathcal{C}_{\text{punc}}) \le k\).
If \(s < d\), suppose that a nonzero codeword \(M \in \mathcal{C}\) satisfies \(M|_{\overline{J}} = 0\). Then all nonzero entries of \(M\) lie in \(s\) columns, so \(\operatorname{rank}(M) \le s\). Since \(M\) is nonzero, \(\operatorname{rank}(M) \ge d\), contradicting \(s < d\). Thus the kernel is \(\{0\}\), and the map $\Pi|_{\mathcal{C}}$ is injective and preserves dimension.

(2) Deleting \(s\) columns decreases rank by at most \(s\). Thus \(\operatorname{rank}(M|_{\overline{J}}) \ge \operatorname{rank}(M) - s\) for every \(M \in \mathcal{C}\), and consequently \(d(\mathcal{C}_{\text{punc}}) \ge d - s\).
\end{proof}

\begin{lemma}\cite{Neri2019}\label{lem:shorten}
Let $\mathcal{C}$ be an $[n\times l,k,d]_q$ rank-metric code, and let \(J \subseteq [l]\) with \(s = |J|\). Then
	\begin{enumerate}
		\item \(\dim(\mathcal{C}_{\text{short}}) \ge k - n s\).
		\item \(d(\mathcal{C}_{\text{short}}) \ge d\).
	\end{enumerate}
\end{lemma}

\begin{proof}
(1) Requiring the \(s\) columns indexed by \(J\) to be zero imposes \(n \times s\) linear constraints. Each independent constraint decreases the dimension by one, so the dimension decreases by at most \(ns\).

(2) Let \(M'\) be an arbitrary nonzero codeword of \(\mathcal{C}_{\text{short}}\), corresponding to a matrix \(M\) in the original code. Since \(M\) has nonzero columns only in \(\overline{J}\) and is zero in the columns indexed by \(J\), the ranks of \(M\) and \(M'\) are equal, namely \(\operatorname{rank}(M) = \operatorname{rank}(M')\).
Since \(M \in \mathcal{C}\) is nonzero, \(\operatorname{rank}(M) \ge d\). Thus \(\operatorname{rank}(M') \ge d\), and hence \(d(\mathcal{C}_{\text{short}}) \ge d\).
\end{proof}

We conclude this section with a lemma that will be used repeatedly in the subsequent constructions.

\begin{lemma}\label{vmin}
Let $\mathcal D =[c_1,c_2,\cdots,c_l]$ be a Ferrers diagram, and let $1\le d \le d' \le l$. If there exists an $[\mathcal D, \nu_0(\mathcal D, d), d']_q$ code $\mathcal C$, then $\mathcal C$ is an optimal $[\mathcal D, \nu_0(\mathcal D, d), d]_q$ code.
\end{lemma}
\begin{proof}
Since $\dim \mathcal{C}=\nu_0(\mathcal D, d)$, we have $\nu_0(\mathcal D, d)=\dim \mathcal{C}\le \nu_{min}(\mathcal D, d')$. If $\mathcal{C}={0}$, then the conclusion holds trivially. Otherwise, suppose that $d'> d$. Then $\nu_{min}(\mathcal D, d') < \nu_{min}(\mathcal D, d) \le \nu_0(\mathcal D, d)$, which yields a contradiction. Hence $d'=d$, and $\mathcal{C}$ is an optimal $[\mathcal D, \nu_0(\mathcal D, d), d]_q$ code.
\end{proof}

\section{Main constructions}\label{sec:construction}

\subsection{The chain of kernels of a nilpotent map}
Let $p$ be a prime number, $q$ and $P$ be powers of $p$, and let $e$ be a positive integer. Set $n=eP$.
Let $F=\mathbb F_q$, $E=\mathbb F_{q^e}$ and $L=\mathbb F_{q^n}$. Let \(\sigma\) be the Frobenius $F$-automorphism of \(L\), given by $x\mapsto x^q$. Since $\sigma^e(x)=x^{q^e}=x$ for every $x\in E$, the map $\sigma^e$ is the Frobenius $E$-automorphism of \(L\). Put $$N=\sigma^e-\mathrm{id}.$$ Then $N$ is an $E$-linear endomorphism of $L$ satisfying
$N^P=0$ and $N^{P-1}\neq 0$.
Thus there exists \(w\in L\) such that
the vectors
$$\{w, N(w), N^2(w), \dots, N^{P-1}(w)\}$$
form an \(E\)-basis of \(L\). For $1\leq i\leq P$, let $v_i=N^{i-1}(w)$. The subspace chain
	\[
\{0\} = V_0 \subsetneq V_1 \subsetneq V_2 \subsetneq \dots \subsetneq V_{P-1}\subsetneq V_P= L
	\]
 induced by this basis is a complete flag of \(L\) as an \(E\)-vector space, where \(V_i = \ker N^i = \operatorname{span}_E(v_{P-i+1}, \dots, v_{P})\).

Throughout this section, we retain the notation introduced above unless otherwise specified. In the lemma below, part (1) is an elementary algebraic fact, while part (2) is taken from \cite[Proposition 3.3]{neri2024}; these properties play a key role in the proof of our main theorem.

\begin{lemma}{\cite[Proposition~3.3]{neri2024}}\label{kersuit} Let $1\leq i,j\leq P$. Then
	\begin{itemize}
	\item[(1)]  $\sigma (V_i)\subseteq V_i$, $N^j(V_i)\subseteq V_{\max\{0,i-j\}}.$
	\item[(2)] $ V_iV_j\subseteq V_{\min\{i+j-1,P\}}$.
\end{itemize}
	\end{lemma}
	\begin{proof}
We include the proof for completeness.

(1) Let $x\in V_i$. Then $N^i(x)=0$, so $N^i(\sigma(x))=\sigma(N^i(x))=\sigma(0)=0$, and therefore $\sigma(x)\in V_i$.
If $i\leq j$, then $N^j(V_i)=\{0\}=V_0$. If $i> j$, then
$N^{i-j}(N^j(x))=N^i(x)={0}$, so $N^j(x)\in V_{i-j}$.
Therefore $N^j(V_i)\subseteq V_{\max\{0,i-j\}}$.

(2) Put $\tau =\sigma^e$.
Since $v_P=N^{P-1}(w)$, we have $N(v_P)=0$, and hence $\sigma^e(v_P)=v_P^{q^e}=v_P$. Thus $v_P\in E$.
Consequently, $V_1=\mathrm{span}(v_P)=E$.
For $i=j=1$, we have $V_iV_j=E=V_1$, so the assertion holds. Let $k$ be a positive integer, and assume that $V_iV_j\subseteq V_{\min\{i+j-1,P\}}$ whenever $i+j\leq k$. Let $x\in V_i$, $y\in V_j$ and $i+j=k+1$. Put $u=Nx$ and $v=Ny$. By~(1), $u\in V_{i-1}$ and $v\in V_{j-1}$. Then
	\[
	\begin{aligned}
		N(xy)&=\tau(xy)-xy=\tau(x)\tau(y)-xy\\
		&=(Nx+x)(Ny+y)-xy=xv+yu+uv.
	\end{aligned}
	\]
The right-hand side belongs to $V_iV_{j-1}+V_jV_{i-1}+V_{i-1}V_{j-1}$. By the induction hypothesis, $N(xy)\in V_{i+j-2}$, so $xy\in V_{i+j-1}$. This proves~(2).
	\end{proof}

The next proposition gives an $L$-basis of $L[\sigma]$  and provides the starting point for our subsequent constructions of Ferrers diagram rank-metric codes.

	\begin{proposition}\label{prop:mixed-basis}
For \(0\le i<P\) and \(0\le r<e\), put
		$	v_{i,r}=\sigma^rN^i$.
Then $$\mathscr C=\{v_{i,r}:\ 0\le i<P,\ 0\le r<e\}$$ is an $L$-basis of $L[\sigma]$.

	\end{proposition}

	\begin{proof}
It suffices to prove that the coordinate matrix of $\{v _{i,r}:\ 0\le i< P,\ 0\le r< e\}$ with respect to $\{\mathrm{id},\sigma,\cdots,\sigma^{n-1}\}$ is invertible. Since \(\sigma\) and \(\mathrm{id}\) commute, the binomial theorem gives
		\[
		v_{i,r}=\sigma^rN^i=\sigma^r(\sigma^e-\mathrm{id})^i
		=\sigma^{ei+r}+\sum_{k=0}^{i-1}\binom{i}{k}(-1)^{i-k}\sigma^{ek+r},
		\]
Thus \(v_{i,r}\) is the sum of \(\sigma^{ei+r}\) and terms whose powers of \(\sigma\) are less than \(ei+r\).
Therefore
		\[
		\begin{aligned}
			&(\mathrm{id},v_{0,1},\ldots,v_{0,e-1},\ldots,v_{P-1,1},\ldots,v_{P-1,e-1})\\
			&=(\mathrm{id},\sigma,\cdots,\sigma^{n-1})
			\begin{pmatrix}
				1&*&*&\cdots&*\\
				0&1&*&\cdots&*\\
				0&0&1&\cdots&*\\
				\vdots&\vdots&\vdots&\ddots&*\\
				0&0&0&\cdots&1
			\end{pmatrix}
		\end{aligned}
		\]
The matrix on the right is invertible, so $\mathscr C$ is an $L$-basis of $L[\sigma]$.
	\end{proof}

\subsection{Optimal Ferrers diagram rank-metric codes for block monotone diagrams}

To construct rank-metric codes over $F$ using the extension field $L$, we choose a basis of $L$ as an $F$-vector space.
Let ${u_1,u_2,\cdots,u_{e}}$ be an $F$-basis of $E$, and put
$$ \mathscr B=\{u_iv_j\mid 1\leq i\leq e, 1\leq j\leq P\}.$$
Then $\mathscr B$ is an $F$-basis of $L$.

	\begin{proposition}\label{FerrRMC}
Let \(k_1,k_2,\cdots,k_P\) be integers with \(1\leq  k_1<k_2<\cdots<k_w= k_{w+1}=\cdots=k_P=P\). Let
\[\mathcal D=\big[[k_1,k_2,\cdots,k_w,\cdots,k_P]\big]_e.\] Put
		\[
		S= \left\{ \sum_{i=1}^{P} \sum_{r=0}^{e-1} \lambda_{i,r} \sigma^r N^{i-1}\mid \lambda_{i,r} \in V_{k_{i}} \right\}.
		\]
Then
		$
		\phi_{\mathscr{B}}(S) = F^{\mathcal{D}}.
		$
	\end{proposition}

	\begin{proof}
Order $\mathscr B$ as follows:
$$\mathscr B=(u_1v_P,u_2v_P,\cdots,u_ev_P, u_1v_{P-1},u_2v_{P-1},\cdots,u_ev_{P-1},\cdots,u_1v_1,u_2v_1,\cdots,u_ev_1).$$
Let $f=\lambda_{i,r} \sigma^r N^{i-1} \in S$, where $\lambda_{i,r} \in V_{k_i}$. For every $1\le j\le P$, we have $$V_j=\mathrm{span}_F\{u_iv_s\mid1\le i\le e,\ P-j+1\le s\le P\}.$$ 
To prove $
\phi_{\mathscr{B}}(S) = F^{\mathcal{D}}
$, it suffices to show that
 $f(V_j)\subseteq V_{k_j}$ for every $j$.
\begin{itemize}
\item[(i)] Suppose that $i-1\geq j$. By Lemma~\ref{kersuit}, $N^{i-1} (V_j)=0$, so $f(V_j)=\lambda_{i,r} \sigma^r N^{i-1} (V_j)=0\subseteq V_{k_j}$.
\item[(ii)] Suppose that $i-1< j$.
If $j\geq w$, then $V_{k_j}=L$, and clearly $f(V_j)\subseteq V_{k_j}$.
Assume that $j< w$.
By Lemma~\ref{kersuit},
$$f(V_j)=\lambda_{i,r}\sigma^rN^{i-1}(V_j)\subseteq V_{k_i}V_{j-i+1}\subseteq V_{k_i+j-i+1-1}=V_{k_i+j-i}.$$
Since $(k_1, k_2,\cdots,k_w)$ is strictly increasing, $k_j-k_i\geq j-i$. Hence $k_i+j-i\leq k_j$, and therefore $f(V_j)\subseteq V_{k_i+j-i}\subseteq V_{k_j}$.
\end{itemize}
This proves the proposition.
\end{proof}
In what follows, we state our main construction theorem. We first consider $e$-block monotone Ferrers diagrams with $n=eP$ rows.
\begin{theorem}\label{mainthm1}
Let \(k_1,k_2,\cdots,k_s\) be integers such that \(1\leq k_1<k_2<\cdots<k_w= k_{w+1}=\cdots=k_s=P, s\le P\). Let \[\mathcal D=\big[[k_1,k_2,\cdots,k_w,\cdots,k_s]\big]_e.\] For every $1\leq d\leq se$, there exists an optimal \([\mathcal{D}, \nu_{0}(\mathcal{D}, d), d]_q\) code. That is, there exist optimal codes over finite fields of characteristic $p$ for $e$-block monotone Ferrers diagrams whose rightmost column has $n=Pe$ dots.
\end{theorem}
\begin{proof}
First suppose that $s=P$, so \[\mathcal D=\big[[k_1,k_2,\cdots,k_w,\cdots,k_P]\big]_e.\] Let $1\leq d\leq Pe$ and 
\[
S_{d}= \left\{ \sum_{i=1}^{P} \sum_{r=0}^{e-1} \lambda_{i,r} \sigma^r N^{i-1}\mid e(i-1)+r\leq n-d,\ \lambda_{i,r} \in V_{k_{i}} \right\}.
\]
Let \(\mathcal C_{\mathcal D,d} = \phi_{\mathscr{B}}(S_{d})\), where $\mathscr{B}$ is the ordered basis in the proof of Proposition \ref{FerrRMC}. We prove that $C_{\mathcal D,d}$ is an optimal \([\mathcal{D}, \nu_{0}(\mathcal{D}, d), d]_q\) code.
Let $0\neq f\in S_{d}$.
Since $\deg_\sigma (f)\leq n-d$, Lemma~\ref{lem:degree} gives $\operatorname{rank}\phi_{\mathscr B}(f)\geq d$. Since $\phi_{\mathscr B}$ is an isomorphism, $\dim  \mathcal C_{\mathcal D,d} = \dim \phi_{\mathscr{B}}(S_{d})=v_0(\mathcal{D}, d)$, which is the total number of dots in the first $n-d+1$ columns of $\mathcal{D}$.
By Proposition~\ref{FerrRMC}, $
\phi_{\mathscr{B}}(S_d) \subseteq F^{\mathcal{D}}
$. Hence $\mathcal C_{\mathcal D,d}$ is an optimal \([\mathcal{D}, \nu_{0}(\mathcal{D}, d), d]_q\) code.

If $s<P$, then \[\mathcal D=\big[[k_1,k_2,\cdots,k_w,\cdots,k_s]\big]_e.\] Let $$\mathcal G=\big[[k_1,k_2,\cdots,k_w,\cdots,k_s,\underset{(P-s)\ \text{times}}{\underbrace{P,\cdots,\cdots,P}}]\big]_e.$$
Put $d'=(P-s)e$, and let $J\subseteq [eP]$ with $\overline{J}=[se]$.
By the case already proved, there exists an optimal $[\mathcal G,\nu_{0}(\mathcal G,d+d'),d+d']_q$ code $\mathcal C$. By Lemma~\ref{lem:puncture}, puncturing $\mathcal C$ at the columns indexed by $[eP]\setminus [se]$ gives
an $[\mathcal G|_{\overline J},\nu_{0}(\mathcal G,d+d'),d'']_q$ code $\mathcal C|_{\overline{J}}$, where $\mathcal G|_{\overline J}=\mathcal D$ and $d''\geq d$. A direct computation gives $v_0(\mathcal G,d+d')=v_0(\mathcal D,d)$.
By Lemma~\ref{vmin},
$\mathcal C|_{\overline{J}}$ is an optimal $[\mathcal D,\nu_{0}(\mathcal D,d),d]_q$ code.
\end{proof}

\begin{example}\label{ex3_1} Let $\mathcal D=\big[[2,3,4,4]\big]_3$ be a Ferrers diagram:
	$$\renewcommand{\arraystretch}{0.8}
	\begin{array}{c@{\quad}c@{\quad}c@{\quad}c@{\quad}c@{\quad}c@{\quad}c@{\quad}c@{\quad}c@{\quad}c@{\quad}c@{\quad}c}

		\bullet & \bullet & \bullet & \bullet & \bullet & \bullet & \bullet & \bullet & \bullet & \bullet & \bullet & \bullet \\
		\bullet & \bullet & \bullet & \bullet & \bullet & \bullet & \bullet & \bullet & \bullet & \bullet & \bullet & \bullet \\
		\bullet & \bullet & \bullet & \bullet & \bullet & \bullet & \bullet & \bullet & \bullet & \bullet & \bullet & \bullet \\

		\bullet & \bullet & \bullet& \bullet & \bullet & \bullet & \bullet & \bullet & \bullet & \bullet & \bullet & \bullet \\
		\bullet & \bullet & \bullet & \bullet & \bullet & \bullet & \bullet & \bullet & \bullet & \bullet & \bullet & \bullet \\
		\bullet & \bullet & \bullet & \bullet & \bullet & \bullet & \bullet & \bullet & \bullet & \bullet & \bullet & \bullet \\

		& & &\bullet  &\bullet  & \bullet & \bullet & \bullet & \bullet & \bullet & \bullet & \bullet \\
		& & &\bullet  & \bullet & \bullet & \bullet & \bullet & \bullet & \bullet & \bullet & \bullet \\
		& & & \bullet & \bullet & \bullet & \bullet & \bullet & \bullet & \bullet & \bullet & \bullet \\

		& & & & & &\bullet  & \bullet &\bullet  & \bullet & \bullet & \bullet \\
		& & & & & & \bullet & \bullet & \bullet & \bullet & \bullet & \bullet \\
		& & & & & & \bullet &\bullet  & \bullet & \bullet & \bullet & \bullet \\
	\end{array}$$
Take $p=q=2$, $e=3$, $P=4$ and $n=eP=12$. For $2\leq d\leq 12$, Theorem~
\ref{mainthm1} gives an optimal $[\mathcal D,\nu_0(\mathcal D,d),d]_2$ code.
For example, if $d=9$, then $\nu_0(\mathcal D,9)=27$. We construct an optimal $[\mathcal D,27,9]_2$ code.
Let $F=\mathbb F_2$, $E=\mathbb F_8=F(\omega)$ and $L=\mathbb F_{2^{12}}=F(\alpha)$, where $\alpha$ is a root of the primitive polynomial $x^{12}+x^6+x^4+x+1$.

Take $\omega=\alpha^{585}$. Then
\[
\omega^3=\omega+1,\qquad
\omega=\alpha^9+\alpha^8+\alpha^7+\alpha^5+\alpha^3+1.
\]
Define $\sigma(x)=x^2$ and $N=\sigma^3-\mathrm{id}$.
Take $w_4=\alpha$ and $w_i=N^{4-i}(\alpha)$ for $1\le i\le3$. Then
\[
\begin{aligned}
w_1&=\omega, & w_2&=\alpha^7+\alpha^6+\alpha^2,\\
w_3&=\alpha^8+\alpha, & w_4&=\alpha.
\end{aligned}
\]
Since $N^4=0$ and $N^3(\alpha)=\omega\ne0$,
$(w_1,w_2,w_3,w_4)$ is an $E$-basis of $L$, and
\[
V_i=\ker N^i=\operatorname{span}_E\{w_1,\ldots,w_i\},
\qquad 1\le i\le4.
\]
Choose the following ordered $F$-basis of $L$:
\[
\mathcal B=(w_1,w_1\omega,w_1\omega^2,
 w_2,w_2\omega,w_2\omega^2,
 w_3,w_3\omega,w_3\omega^2,
 w_4,w_4\omega,w_4\omega^2).
\]

Since $n-d=3$, consider the $F$-linear space
\[
S_9=\left\{
\lambda_0\mathrm{id}+\lambda_1\sigma+\lambda_2\sigma^2+\lambda_3N:
\lambda_0,\lambda_1,\lambda_2\in V_2,\ \lambda_3\in V_3
\right\}.
\]
\(\dim_F S_9=6\times3+9=27\). An \(F\)-basis of $S_9$ is
\[
\{w_i\omega^j\sigma^r:0\le r\le2,\ 1\le i\le2,\ 0\le j\le2\}
\ \cup\
\{w_i\omega^jN:1\le i\le3,\ 0\le j\le2\}.
\]
Their representation matrices $M_1,\dots,M_{27}$ with respect to $\mathcal B$ (see Appendix~A)
generate the \(\mathbb F_2\)-linear rank-metric code
\[
\mathcal C_{\mathcal D,9}
=\phi_{\mathcal B}(S_9)
=\left\{\sum_{j=1}^{27}a_jM_j:a_j\in F\right\}.
\]
This code has $n=12$ rows and dimension $k=27$.
The computed rank distribution of its $2^{27}-1=134\,217\,727$ nonzero codewords is given in Table 1.

\begin{table}[H]
\centering
\caption{The rank distribution of $\mathcal C_{\mathcal D,9}$}
\begin{tabular}{c@{\qquad}rrrr}
\toprule
Rank & $9$ & $10$ & $11$ & $12$\\
\midrule
Number of codewords & $3\,158\,589$ & $21\,773\,738$ & $74\,045\,272$ & $35\,240\,128$\\
\bottomrule
\end{tabular}
\end{table}

Thus the minimum rank distance of this code is \(d(\mathcal C_{\mathcal D,9})=9\). Therefore $\mathcal C_{\mathcal D,9}$ is an optimal $[\mathcal D,27,9]_2$ code.
\end{example}

\begin{remark}
In \cite[Theorem~4.2]{neri2024}, optimal Ferrers diagram rank-metric codes are constructed over fields of characteristic $p$ for $p^h$-block monotone Ferrers diagrams, where the block size is a power $p^h$ of the characteristic and the number $n$ of rows is also required to be a power of \(p\). Theorem~\ref{mainthm1} relaxes the restrictions on both the block size and the number $n$ of rows, extending that result. For instance, the optimal Ferrers
diagram rank-metric code in Example~\ref{ex3_1} cannot be obtained by \cite[Theorem~4.2]{neri2024}.
\end{remark}
We now prove that, for $e$-block strictly monotone Ferrers diagrams, the restriction $n=eP$ on the number of rows can be removed.
\begin{theorem} \label{mainthm2}
Let \(k_1,k_2,\cdots,k_s\) be integers with \(1\leq k_1<k_{2}<\cdots<k_s\). Let \[\mathcal D=\big[[k_1,k_2,\cdots,k_s]\big]_e.\]
For every $1\leq d\leq se$, there exists an optimal \([\mathcal{D}, \nu_{0}(\mathcal{D}, d), d]_q\) code. Thus optimal Ferrers diagram rank-metric codes exist over every finite field $F$ for $e$-block strictly monotone Ferrers diagrams. In particular, optimal codes exist for $e$-block upper triangular Ferrers diagrams.

\end{theorem}
\begin{proof}
If $k_s$ is a power of the characteristic $p$, then by Theorem\ref{mainthm1} there exists an optimal $[\mathcal D,\nu_{0}(\mathcal D,d),d]_q$ code. 
If $k_s$ is not a power of the characteristic $p$, 
let $t$ be the smallest positive integer such that $p^te>k_se$. Set $P=p^t$ and $n'=eP$.
Let $$\mathcal G=\big[[k_1,k_2,\cdots,k_s,\underset{(P-s)\ \text{times}}{\underbrace{P,\cdots,\cdots,P}}]\big]_e.$$
Put $d'=(P-s)e$, and let $J\subseteq [n']$ with $\overline{J}=[se]$.
By Theorem~\ref{mainthm1}, there exists an optimal $[\mathcal G,\nu_{0}(\mathcal G,d+d'),d+d']_q$ code $\mathcal C$. By Lemma~\ref{lem:puncture}, puncturing $\mathcal C$ at the columns indexed by $[n']\setminus [se]$ gives
an $[\mathcal G|_{\overline J},\nu_{0}(\mathcal G,d+d'),d'']_q$ code $\mathcal C|_{\overline{J}}$, where $\mathcal G|_{\overline J}=\mathcal D$ and $d''\geq d$. A direct computation gives $v_0(\mathcal G,d+d')=v_0(\mathcal D,d)$.
By Lemma~\ref{vmin},
$\mathcal C|_{\overline{J}}$ is an optimal $[\mathcal D,\nu_{0}(\mathcal D,d),d]_q$ code.
\end{proof}

\begin{remark}
In \cite[Theorem~4.9]{neri2024}, optimal codes are constructed for $p^h$-block strictly monotone diagrams over finite fields of characteristic $p$.
In \cite[Theorem~5.9]{calderini2026}, optimal codes are constructed for $se\times se$ $e$-block upper triangular Ferrers diagrams over any finite field when $d\in\{1,2,\cdots, e+1\}\cup\{se-1,se\}$.
Theorem~\ref{mainthm2} removes the restrictions on the characteristic and minimum distance, extending these results.
\end{remark}
\begin{remark}
Xu and Li \cite{xu2025} proposed a new approach to selecting skeleton codes based on one-factorizations of complete graphs and used the multilevel construction to obtain $(nt,M,4t,3t)_q$ CDCs for $n=(m+3)\ge 8$ and even $t$. By computing the dimensions of the associated Ferrers diagram rank-metric codes, they improved the lower bounds on $A_q(n,8,6)$ for $16\le n\le 19$. These dimension computations rely on existing Ferrers diagram rank-metric codes. In \cite[Sec.3.2.1]{xu2025}, the case of $\mathcal D=[\underbrace{t,\dots,t}_{t},\underbrace{2t,\dots,2t}_{t},\underbrace{3t,\dots,3t}_{t}]$ with distance $d=2t$ is considered.
Since the existence of the corresponding optimal Ferrers diagram rank-metric codes was unknown when $m=5,t\ne2$ and $q<4$, they constructed Ferrers diagram rank-metric codes of dimension $3t$. Theorem~\ref{mainthm2} now gives the corresponding optimal Ferrers diagram rank-metric codes, increasing the dimension from $3t$ to $t^2+2t$. 
\end{remark}

\begin{example}\label{ex3_2}
Let $\mathcal D=\big[[1,2,3,4]\big]_3$ be a Ferrers diagram: 
		$$\renewcommand{\arraystretch}{0.8}
	\begin{array}{c@{\quad}c@{\quad}c@{\quad}c@{\quad}c@{\quad}c@{\quad}c@{\quad}c@{\quad}c@{\quad}c@{\quad}c@{\quad}c}

		\bullet & \bullet & \bullet & \bullet & \bullet & \bullet & \bullet & \bullet & \bullet & \bullet & \bullet & \bullet \\
		\bullet & \bullet & \bullet & \bullet & \bullet & \bullet & \bullet & \bullet & \bullet & \bullet & \bullet & \bullet \\
		\bullet & \bullet & \bullet & \bullet & \bullet & \bullet & \bullet & \bullet & \bullet & \bullet & \bullet & \bullet \\

		& & & \bullet & \bullet & \bullet & \bullet & \bullet & \bullet & \bullet & \bullet & \bullet \\
		& & & \bullet & \bullet & \bullet & \bullet & \bullet & \bullet & \bullet & \bullet & \bullet \\
		& & & \bullet & \bullet & \bullet & \bullet & \bullet & \bullet & \bullet & \bullet & \bullet \\

		& & & & & & \bullet & \bullet & \bullet & \bullet & \bullet & \bullet \\
		& & & & & & \bullet & \bullet & \bullet & \bullet & \bullet & \bullet \\
		& & & & & & \bullet & \bullet & \bullet & \bullet & \bullet & \bullet \\

		& & & & & & & & & \bullet & \bullet & \bullet \\
		& & & & & & & & & \bullet & \bullet & \bullet \\
		& & & & & & & & & \bullet & \bullet & \bullet \\
	\end{array}$$

\noindent Let $d=10$. Then $\nu_{\min}(\mathcal D,10)=9$.
We construct an optimal $[\mathcal D,9,10]_2$ code.
Let
\(F=\mathbb F_2\) and \(L=\mathbb F_{2^{12}}=\mathbb F_2(\alpha)\), where \(\alpha\) is a root of the primitive polynomial \(x^{12}+x^6+x^4+x+1\). Put \(\omega=\alpha^{2340}\) and \(E=\mathbb F_2(\omega)\), then \(\omega^3=\omega+1\). Thus \(q=p=2\), \(e=[E:F]=3\) and \(P=[L:E]=4\). 
Define the Frobenius automorphism \(\sigma(x)=x^2\) and the nilpotent operator \(N=\sigma^3-\mathrm{id}\). Then \(N^4=0\) and \(N^3\neq0\). Take \(w_4=\alpha\) and put \(w_i=N^{4-i}(w_4)\) for \(1\le i\le 4\). This gives an \(E\)-basis \(\{w_1,\dots,w_4\}\) of \(L\), inducing the subspace chain
	\[
	0=V_0\subsetneq V_1\subsetneq V_2\subsetneq V_3\subsetneq V_4=L,
	\]
where \(V_1=E\). Moreover, choose the following ordered $F$-basis of $L$:
\[
\mathcal B=(w_1,w_1\omega,w_1\omega^2,
 w_2,w_2\omega,w_2\omega^2,
 w_3,w_3\omega,w_3\omega^2,
 w_4,w_4\omega,w_4\omega^2).
\]

Since $n-d=2$, consider the \(F\)-linear space
	\[
	S_{10}=\left\{\lambda_{0}\mathrm{id}+\lambda_1\sigma+\lambda_2\sigma^2 \ \middle|\ \lambda_{0},\lambda_1,\lambda_2\in V_1=E\right\}.
	\]
      \(\dim_F S_{10}=3\times3=9\). Choose the \(F\)-basis \(\{w_1\omega^j\sigma^r: 0\le j,r\le 2\}\) of $S_{10}$. Their representation matrices \(M_1,\dots,M_9\) with respect to \(\mathscr  B\) (see Appendix~B) generate the \(\mathbb F_2\)-linear rank-metric code
	\[
	\mathcal C_{\mathcal{D},10}=\left\{\sum_{j=1}^{9}a_jM_j \ \middle|\ a_j\in\mathbb F_2\right\}.
	\]
This code has \(n=12\) rows and dimension \(k=9\). The computed rank distribution of its \(2^9-1=511\) nonzero codewords is given in Table 2:

\begin{table}[H]
\centering
\caption{The rank distribution of $\mathcal C_{\mathcal D,10}$}
		\begin{tabular}{cccc}
			\toprule
Rank & \(10\) & \(11\) & \(12\)\\
			\midrule
Number of codewords & \(245\) & \(98\) & \(168\)\\
			\bottomrule
		\end{tabular}
\end{table}
Thus the minimum rank distance of this code is \(d(\mathcal C_{\mathcal{D},10})=10\). Therefore $\mathcal C_{\mathcal{D},10}$ is an optimal $[\mathcal D,9,10]_2$ code.
\end{example}

Calderini, Messia and Neri \cite{calderini2026} used sum-rank-metric codes to construct optimal Ferrers diagram rank-metric codes for $e$-block Ferrers diagrams. They introduced MSRD-constructible block diagram pairs \cite[Definition~3.7]{calderini2026} and proved that optimality for $e$-block upper triangular diagrams implies optimality for MSRD-constructible block diagram pairs \cite[Theorem~5.3, Remark~5.4]{calderini2026}.
As an immediate consequence of Theorem~\ref{mainthm2} and \cite[Theorem~5.3]{calderini2026}, we obtain the following corollary.

\begin{corollary}\label{MSRD_block}
Let $(\mathcal D,d)$ be an MSRD-constructible block diagram pair. Then, over every finite field $ F$, there exists an optimal \([\mathcal{D}, \nu_{\min}(\mathcal{D}, d), d]_q\) code.
\end{corollary}

For block monotone diagrams, we cannot remove the restriction that the last column contains $n=eP$ dots. However, we can prove the existence of optimal Ferrers diagram rank-metric codes when the number of full columns satisfies certain conditions.

	\begin{proposition} \label{prop-mon-block}
Let \(k_1,k_{2},\cdots,k_r,\cdots,k_t\) be integers with \(1\leq k_1<k_{2}<\cdots<k_r= k_{r+1}=\cdots=k_t\). Suppose that $k_{r-1}+(t-r+1)\leq k_t$. For \[\mathcal D=\big[[k_1,k_2,\cdots,k_t]\big]_e\]
and every $1\leq d\leq te$, there exists an optimal \([\mathcal{D}, \nu_{0}(\mathcal{D}, d), d]_q\) code.
\end{proposition}

\begin{proof} If $d\leq (t-r+1)e+1$, then the rightmost $d-1$ columns of $\mathcal D$ are full and Proposition~\ref{fullpoints} gives an optimal \([\mathcal{D}, \nu_{0}(\mathcal{D}, d), d]_q\) code.
Suppose that $d> (t-r+1)e+1$. Since $k_{r-1}+(t-r+1)\leq k_t$, we can define the Ferrers diagram
	$$\mathcal F=\big[[k_1,k_2,\cdots,k_{r-1}, k_{r-1}+1,\cdots, k_{r-1}+(t-r+1)]\big]_e,$$
Then $\mathcal F$ is an $e$-block strictly monotone diagram and $\mathcal F \subseteq \mathcal D$. By Theorem~\ref{mainthm2}, there exists an
optimal $[\mathcal F,\nu_{0}(\mathcal F,d),d]_q$ code, and we have $\nu_{\min}(\mathcal F,d)=\nu_0(\mathcal F,d)$. Since the first $et-d+1$ columns are unchanged, we have $\nu_0(\mathcal D,d)=\nu_0(\mathcal F,d)$. For every $i\in [d-1]$, $\nu_0(\mathcal D,d)=\nu_0(\mathcal F,d)\le \nu_i(\mathcal F,d) \le \nu_i(\mathcal D,d)$, so $\nu_{\min}(\mathcal D,d)=\nu_0(\mathcal D,d)$ and then $\nu_{\min}(\mathcal F,d)=\nu_{\min}(\mathcal D,d)$. By   Lemma~\ref{pendingdot}, there exists an optimal \([\mathcal{D}, \nu_{0}(\mathcal{D}, d), d]_q\) code.
\end{proof}

\subsection{Optimal Ferrers diagram rank-metric codes for quasi-block monotone diagrams}
We can first embed a quasi-block monotone diagram in a larger block monotone diagram and then use puncturing and shortening of rank-metric codes to construct an optimal Ferrers diagram rank-metric code.

We first give an example.
Let $d=6$ and consider $\mathcal{D} = \big[ [3^{(2)}, 6^{(2)}, 9^{(1)}, 12^{(3)}]\big]_3$.
We construct an optimal $[\mathcal{D},12,6]_2$ code.
Complete it to the block upper triangular diagram $\mathcal G=\big[[1,2,3,4]\big]_3$:
\[
\mathcal{D} =
\renewcommand{\arraystretch}{0.8}
\begin{array}{c@{\quad}c@{\quad}c@{\quad}c@{\quad}c@{\quad}c@{\quad}c@{\quad}c}
	\bullet & \bullet & \bullet & \bullet & \bullet & \bullet & \bullet & \bullet \\
	\bullet & \bullet & \bullet & \bullet & \bullet & \bullet & \bullet & \bullet \\
	\bullet & \bullet & \bullet & \bullet & \bullet & \bullet & \bullet & \bullet \\
	& & \bullet & \bullet & \bullet & \bullet & \bullet & \bullet \\
	& & \bullet & \bullet & \bullet & \bullet & \bullet & \bullet \\
	& & \bullet & \bullet & \bullet & \bullet & \bullet & \bullet \\
	& & & & \bullet & \bullet & \bullet & \bullet \\
	& & & & \bullet & \bullet & \bullet & \bullet \\
	& & & & \bullet & \bullet & \bullet & \bullet \\
	& & & & & \bullet & \bullet & \bullet \\
	& & & & & \bullet & \bullet & \bullet \\
	& & & & & \bullet & \bullet & \bullet \\
\end{array}
\begin{tikzpicture}[remember picture, overlay]

	\draw[dashed,red, line width=1.0pt] (-3.1,3.0) -- (-3.1,-2.8);
\end{tikzpicture}
\hskip 2cm 	\mathcal{G} =
\renewcommand{\arraystretch}{0.8}
\begin{array}{c@{\quad}c@{\quad}c@{\quad}c@{\quad}c@{\quad}c@{\quad}c@{\quad}c@{\quad}c@{\quad}c@{\quad}c@{\quad}c}

	\bullet & \bullet & \circ & \bullet & \bullet & \circ & \bullet & \circ & \circ & \bullet & \bullet & \bullet \\
	\bullet & \bullet & \circ & \bullet & \bullet & \circ & \bullet & \circ & \circ & \bullet & \bullet & \bullet \\
	\bullet & \bullet & \circ & \bullet & \bullet & \circ & \bullet & \circ & \circ & \bullet & \bullet & \bullet \\

	& & & \bullet & \bullet & \circ & \bullet & \circ & \circ & \bullet & \bullet & \bullet \\
	& & & \bullet & \bullet & \circ & \bullet & \circ & \circ & \bullet & \bullet & \bullet \\
	& & & \bullet & \bullet & \circ & \bullet & \circ & \circ & \bullet & \bullet & \bullet \\

	& & & & & & \bullet & \circ & \circ & \bullet & \bullet & \bullet \\
	& & & & & & \bullet & \circ & \circ & \bullet & \bullet & \bullet \\
	& & & & & & \bullet & \circ & \circ & \bullet & \bullet & \bullet \\

	& & & & & & & & & \bullet & \bullet & \bullet \\
	& & & & & & & & & \bullet & \bullet & \bullet \\
	& & & & & & & & & \bullet & \bullet & \bullet \\
\end{array}
\begin{tikzpicture}[remember picture, overlay]

	\draw[dashed,red, line width=1.0pt] (-5.0,3.0) -- (-5.0,-2.8);
\end{tikzpicture}
\]

\medskip
\noindent
By Theorem~\ref{mainthm2}, there exists an optimal $[\mathcal G,15,9]_2$ code $\mathcal C$. Let $J_l=\{3\}$ and $J_r=\{6,8,9\}$. Puncturing $\mathcal C$ at $J_r$ and then shortening at $J_l$ yields an optimal $[\mathcal{D},12,6]_2$ code.

We now prove the existence of optimal Ferrers diagram rank-metric codes for quasi-block strictly monotone diagrams.
\begin{theorem}\label{thm-ni}
Let \(e\) be a positive integer, let \(s_1,s_2,\cdots, s_l\) be positive integers at most \(e\), and let \(k_1<k_2<\cdots <k_l\) be positive integers. Set \(S=\sum_{i=1}^l s_i\), and let
	\[
	\mathcal D= \big[[{(k_1e)}^{(s_1)},(k_2e)^{(s_2)}, \cdots, (k_le)^{(s_l)}]\big]_e.
	\]
Then for every \(1 \le d \le S\), there exists an optimal \([\mathcal{D}, v_{0}(\mathcal{D}, d), d]_q\) code over any finite field.
\end{theorem}

\begin{proof}
First complete \(\mathcal D\) by inserting columns without changing the heights or order of its original
columns to obtain the $e$-block upper triangular diagram
	\[
	\mathcal G=\big[[1,2,\cdots,k_l]\big]_e.
	\]
Let \(J_r\) be the set of indices of the columns added to the right of the \((S-d+1)\)-th column of \(\mathcal D\), and let \(J_l\) be the set of indices of those added to the left of  \((S-d+1)\)-th column of \(\mathcal D\). Let $\overline J_r=[ek_l]\setminus J_r$.

Suppose that $|J_r|=m$. By Theorem~\ref{mainthm2}, there exists an optimal \([\mathcal G, \nu_{0}(\mathcal G,m+d), m+d]_q\) code \(\mathcal C\).

\begin{itemize}
\item By Lemma~\ref{lem:puncture}, puncturing $\mathcal C$ at the columns indexed by ${J_r}$ gives an $[\mathcal G|_{\overline J_r}, \nu_{0}(\mathcal G,m+d),d']_q$ code $\mathcal C|_{\overline J_r}$, where $d'\geq d$. Since $\nu_{0}(\mathcal G|_{\overline J_r},d)=\nu_{0}(\mathcal G,m+d)$,
Lemma~\ref{singleton-bound} gives $d'=d$.
\item Let $\mathcal D_1= \mathcal G|_{\overline J_r}$, and let $\mathcal C_1$ be the shortened code of $\mathcal C|_{\overline J_r}$ at $J_l$.
By Lemma~\ref{lem:shorten}, $\mathcal C_1$ is an $[\mathcal D, k, d]_q$ code with $k\geq \nu_{0}(\mathcal G,m+d)-s$, where $s$ is the total number of dots in the columns indexed by $J_l$.
	 \end{itemize}
Finally, $\nu_{0}(\mathcal D,d)= \nu_{0}(\mathcal G,m+d)-s$. Hence $k \ge \nu_{0}(\mathcal D,d)$. Conversely, Lemma~\ref{singleton-bound} gives $k \le \nu_{\min}(\mathcal D,d) \le \nu_{0}(\mathcal D,d)$. Thus $k=\nu_{0}(\mathcal D,d)=\nu_{\min}(\mathcal D,d)$, so $\mathcal C_1$ is an optimal $[\mathcal D, \nu_{0}(\mathcal D,d), d]_q$ code.
\end{proof}
\begin{remark}
For certain quasi-block monotone diagrams, an argument analogous to the proof of Theorem~\ref{thm-ni} yields optimal codes from those of block monotone diagrams.
Let $\mathcal G=\big[[k_1,k_2,\cdots,k_s]\big]_e$ be an $e$-block monotone diagram in Theorem~\ref{mainthm1} or Proposition~\ref{prop-mon-block}. Then, for every $J\subseteq[es]$ and $1\le d'\le es-|J|$, there exists an optimal $[\mathcal G|_{\overline J},\nu_{0}( \mathcal G|_{\overline J},d'),d']_q$ code, where $G|_{\overline J}$ is a quasi $e$-block monotone diagram. To obtain the optimal code, we complete $\mathcal G|_{\overline J}$ to $\mathcal G$ and then find $J_l$ and $J_r$. Next we construct an optimal $[\mathcal G, \nu_{0}(\mathcal G,d'+|J_r|), d'+|J_r|]_q$ code $\mathcal C$. Subsequently, we puncture $\mathcal C$ at $J_r$ and then  shorten the resulting code at $J_l$ to obtain the optimal code.  
\end{remark}
Lemma~\ref{pendingdot} and Theorem~\ref{thm-ni} give the following result.

\begin{proposition}\label{prop-ni-subcode}
Let \(e\) be a positive integer, let \(s_1, s_2, \cdots, s_w\) be positive integers at most \(e\), and let \(k_1 < k_2 < \cdots < k_w\) be positive integers. Put \(S = \sum_{i=1}^w s_i\). For a fixed integer \(1 \le d \le S\), let \(I\) and \(t\) be the unique integers satisfying \(S - d + 1 = \sum_{i=0}^{I-1} s_i + t\), where \(s_0=0\), \(1\le I \le w\) and \(0 \leq t \le s_I\).
Define the Ferrers diagram
		\begin{align*}
			\mathcal{D}= \big[ &
\underbrace{s_{1,1},\dots,s_{1,s_1}}_{s_1\text{ columns}},\dots,
\underbrace{s_{I-1,1},\dots,s_{I-1,s_{I-1}}}_{s_{I-1}\text{ columns}},\
\underbrace{s_{I,1},\dots,s_{I,t}}_{t\text{ columns}},\
\underbrace{k_Ie,\dots,k_Ie}_{s_I-t\text{ columns}},\\
&\underbrace{k_{I+1}e,\dots,k_{I+1}e}_{s_{I+1}\text{ columns}},\dots,
\underbrace{k_we,\dots,k_we}_{s_w\text{ columns}}\big],
		\end{align*}
where
\(s_{i,j} \le k_i e\) for \(1 \le i \le I-1\) and \(1 \le j \le s_i\), and
\(s_{I,j} \le k_I e\) for \(1 \le j \le t\).
Then there exists an optimal \([\mathcal{D}, \nu_{0}(\mathcal D,d), d]_q\) code for every prime power $q$.

\end{proposition}

\begin{remark}
In \cite[Theorem~5.6]{pratihar2023} and \cite[Theorem~5.7]{pratihar2023}, optimal Ferrers diagram rank-metric codes over \(\mathbb F_q\) are constructed for the Ferrers diagrams in Theorem~\ref{thm-ni} and Proposition~\ref{prop-ni-subcode}, respectively, under the assumption that \(q> k_w\). We remove this restriction on $q$, thereby generalizing these results.
\end{remark}
The following example illustrates the construction in Proposition~\ref{prop-ni-subcode}.
\begin{example}
Let $\mathcal D=[2,2,4,4,6,9,9]$ and $d=4$. Embed $\mathcal D$ in the quasi $3$-block strictly monotone diagram $\mathcal G=\big[[3^{(2)},6^{(3)},9^{(2)}]\big]_3$:
\[
\mathcal{D} =
\renewcommand{\arraystretch}{0.8}
\begin{array}{c@{\quad}c@{\quad}c@{\quad}c@{\quad}c@{\quad}c@{\quad}c}
	\bullet &	\bullet & \bullet & \bullet & \bullet & \bullet & \bullet \\
	\bullet &	\bullet & \bullet & \bullet & \bullet & \bullet & \bullet \\
	 &	&  \bullet& \bullet & \bullet & \bullet & \bullet \\
&	& \bullet & \bullet & \bullet & \bullet & \bullet \\
&	& & & \bullet & \bullet & \bullet \\
&	& & & \bullet & \bullet & \bullet \\
&	& & & & \bullet& \bullet \\
&	& & & & \bullet & \bullet \\
&	& & & & \bullet & \bullet \\
\end{array}
\begin{tikzpicture}[remember picture, overlay]

	\draw[dashed,red, line width=1.0pt] (-1.85,2.2) -- (-1.85,-1.9);
\end{tikzpicture}
\hskip 2cm
\mathcal{G} =
\renewcommand{\arraystretch}{0.8}
\begin{array}{c@{\quad}c@{\quad}c@{\quad}c@{\quad}c@{\quad}c@{\quad}c}
	\bullet &	\bullet & \bullet & \bullet & \bullet & \bullet & \bullet \\
	\bullet &	\bullet & \bullet & \bullet & \bullet & \bullet & \bullet \\
	\circ&	\circ&  \bullet& \bullet & \bullet & \bullet & \bullet \\
	&	& \bullet & \bullet & \bullet & \bullet & \bullet \\
	&	& \circ&\circ & \bullet & \bullet & \bullet \\
	&	&\circ &\circ & \bullet & \bullet & \bullet \\
	&	& & & & \bullet& \bullet \\
	&	& & & & \bullet & \bullet \\
	&	& & & & \bullet & \bullet \\
\end{array}
\begin{tikzpicture}[remember picture, overlay]

	\draw[dashed,red, line width=1.0pt] (-1.85,2.2) -- (-1.85,-1.9);
\end{tikzpicture}
\]

\noindent
By Theorem~\ref{thm-ni}, there exists an optimal $[\mathcal G,18,4]_q$ code. We have $\nu_{\min} (\mathcal D,4)=\nu_{\min} (\mathcal G,4)-6=12$. Since $\mathcal D\subseteq \mathcal G$, Lemma~\ref{pendingdot} gives an optimal $[\mathcal D,12,4]_q$ code.

\end{example}
Proposition~\ref{prop-ni-subcode} also yields the following corollary.
\begin{corollary}\label{subcons}
Let \(1=t_0<t_1<t_2<\cdots<t_w\le n\) be integers satisfying
\(t_1\mid t_2\mid\cdots\mid t_w\). Let $l$ and $d$ be positive integers such that
\[
t_{w-1}<l\le t_w,
\qquad
l-t_1+1<d\le l.
\]
Let \(\mathcal D=[c_1,c_2,\cdots,c_l]\) be an \(n\times l\) Ferrers diagram satisfying
\begin{itemize}
    \item \(c_{l-d+1}\le t_1\);
    \item \(c_{l-d+2}\ge t_1\);
\item \(c_{jt_1+1}\ge(j+1)t_1\) for $j\in \{1,2,\cdots,\eta \}$ and $jt_1<l$, where $\eta$ is the quotient in the division of $l$ by $t_1$.
\end{itemize}
Then there exists an optimal $[\mathcal D,\nu_{0}(\mathcal D,d),d]_q$ code over every finite field.
\end{corollary}
\begin{proof}
Consider the subdiagram
\begin{align*}
\mathcal{D'}= \big[ &
\underbrace{c_1,\dots,c_{l-d+1},t_1,\cdots,t_1}_{t_1\text{ columns}},
\underbrace{2t_1,\dots,2t_1}_{t_1\text{ columns}},\ \cdots,
\underbrace{jt_1,\dots,jt_1}_{t_1\text{ columns}},\
\underbrace{(j+1)t_1,\dots,(j+1)t_1}_{t_1\text{ columns}},\\ \cdots,
&\underbrace{\eta t_1,\dots,\eta t_1}_{t_1\text{ columns}},
\underbrace{(\eta+1)t_1,\dots,(\eta+1)t_1}_{l-\eta t_1\text{ columns}}\big],
\end{align*}
By Proposition~\ref{prop-ni-subcode}, there exists an optimal $[\mathcal{D'},\nu_{0}(\mathcal D',d),d]_q$ code. Since $\mathcal D'\subseteq \mathcal D$ and the first $l-d+1$ column heights are
unchanged, $\nu_0(\mathcal D,d)=\nu_0(\mathcal D',d)$.  Moreover, $\nu_0(\mathcal D',d)\le \nu_i(\mathcal D',d)\le \nu_{i}(\mathcal D,d)$ for every $0\le i\le d-1$, thus $\nu_{min}(\mathcal D,d)=\nu_{0}(\mathcal D,d)$ and $\nu_{min}(\mathcal D,d)=\nu_{min}(\mathcal D',d)$. By Lemma~\ref{pendingdot} there exists an optimal $[\mathcal{D},\nu_{0}(\mathcal D,d),d]_q$ code.
\end{proof}
\begin{remark}
The condition $c_{t_i+1}\ge t_{i+1}, i=1,2,\cdots,w-1$ in \cite[Theorem~3.2]{zhang2019} is weakened in the above corollary to $c_{jt_1+1}\ge (j+1)t_1$, $j\in \{1,2,\cdots,\eta \}$ and $jt_1<l$. More precisely, write $k_1=1, t_2={k_2}{t_1}, t_3={k_3}{t_1}, \cdots, t_w={k_w}{t_1}$. Then \(k_1\mid k_2\mid\cdots\mid k_w\). The condition $c_{t_i+1}\ge t_{i+1}, i=1,2,\cdots,w-1$ in \cite[Theorem~3.2]{zhang2019} requires, for every $i$, that the corresponding column heights satisfy $c_{{k_i}{t_1}+1}\ge k_{i+1}t_1,\cdots, c_{{k_i}{t_1}+t_1+1}\ge k_{i+1}t_1, \cdots, c_{{k_{i+1}}{t_1}-1}\ge k_{i+1}t_1, c_{{k_{i+1}}{t_1}}\ge k_{i+1}t_1$. Corollary~\ref{subcons} shows that it suffices to require
$c_{{k_i}{t_1}+1}\ge k_{i}t_1+t_1,\cdots, c_{{k_i}{t_1}+t_1+1}\ge k_{i}t_1+2t_1, \cdots, c_{{k_{i+1}}{t_1}-1}\ge k_{i+1}t_1,  c_{{k_{i+1}}{t_1}}\ge k_{i+1}t_1$, and therefore improves that result.
\end{remark}

\section{Conclusions and problems}\label{sec:conclusion}
In this paper, we present new constructions of optimal codes for block diagrams. For \(e\)-block monotone diagrams with \(eP\) rows, where \(P\) is a power of \(p\) and \(e\) is an arbitrary positive integer, Theorem~\ref{mainthm1} constructs optimal codes for every minimum distance, generalizing the corresponding construction in \cite[Theorem~4.2]{neri2024}. By column puncturing, Theorem~\ref{mainthm2} removes the restriction on the number of rows for block strictly monotone diagrams, demonstrating that optimal codes exist over every finite field for any admissible minimum distance. In particular, this holds for block upper triangular diagrams. Together with \cite[Theorem~5.3]{calderini2026}, we establish the existence of optimal codes for all MSRD-constructible block diagram pairs over every finite field (Corollary~\ref{MSRD_block}).
 As an application to constructions of CDCs, by Theorem~\ref{mainthm2} we increase the dimension of a class of Ferrers diagram rank-metric codes used in the multilevel construction of \cite{xu2025} from \(3t\) to \(t^2+2t\). For block monotone diagrams with repeated full columns, Proposition~\ref{prop-mon-block} reduces optimality to that of block strictly monotone subdiagrams under the condition \(k_{r-1}+(t-r+1)\le k_t\). For quasi-block strictly monotone diagrams, Theorem~\ref{thm-ni} proves the existence of optimal codes by embedding them in block strictly monotone diagrams and using the properties of punctured and shortened rank-metric codes. A similar method applies to certain quasi-block monotone diagrams. Proposition~\ref{prop-ni-subcode} further treats diagrams obtained by removing some dots from the initial columns. These results remove the field-size restrictions in \cite[Theorem~5.6]{pratihar2023} and \cite[Theorem~5.7]{pratihar2023}. Corollary~\ref{subcons} also weakens the conditions on the Ferrers diagram in \cite[Theorem~3.2]{zhang2019}, thereby generalizing that result.

We conclude with two open problems.
\paragraph*{Problem 1}
When $k_{r-1}+(t-r+1)>k_t$, Proposition~\ref{prop-mon-block} does not establish the existence of optimal codes. How can one construct optimal codes for such diagrams? For example, does there exist an optimal Ferrers diagram rank-metric code for $\mathcal D=\big[[2,3,4,4]\big]_3$ and $d=9$ when $q>3$ is an odd prime power?
\paragraph*{Problem 2}
The Jordan canonical form over the base field \(F\) of the nilpotent operator \(N=\sigma^e-\mathrm{id}\) considered here consists of \(e\) nilpotent Jordan blocks, each of size \(P\). Can \(F\)-linear nilpotent operators with unequal Jordan block lengths be considered to establish multiplicative inclusions for their kernel chains? Furthermore, can these properties be used to construct optimal Ferrers diagram rank-metric codes not covered by our results?

\appendix
\section{Basis matrices for Example~\ref{ex3_1}}\label{3_1matrix}
\par\medskip
\noindent
\begin{minipage}[t]{0.32\textwidth}
\centering
\(M_{1}\)\par\smallskip
{\small\renewcommand{\arraystretch}{0.46}\setlength{\arraycolsep}{1.5pt}
\(\left(\begin{array}{ccc|ccc|ccc|ccc}
0 & 0 & 1 & 0 & 0 & 0 & 0 & 0 & 0 & 0 & 0 & 0\\
1 & 0 & 1 & 0 & 0 & 0 & 0 & 0 & 0 & 0 & 0 & 0\\
0 & 1 & 0 & 0 & 0 & 0 & 0 & 0 & 0 & 0 & 0 & 0\\\hline
0 & 0 & 0 & 0 & 0 & 1 & 0 & 0 & 0 & 0 & 0 & 0\\
0 & 0 & 0 & 1 & 0 & 1 & 0 & 0 & 0 & 0 & 0 & 0\\
0 & 0 & 0 & 0 & 1 & 0 & 0 & 0 & 0 & 0 & 0 & 0\\\hline
0 & 0 & 0 & 0 & 0 & 0 & 0 & 0 & 1 & 0 & 0 & 0\\
0 & 0 & 0 & 0 & 0 & 0 & 1 & 0 & 1 & 0 & 0 & 0\\
0 & 0 & 0 & 0 & 0 & 0 & 0 & 1 & 0 & 0 & 0 & 0\\\hline
0 & 0 & 0 & 0 & 0 & 0 & 0 & 0 & 0 & 0 & 0 & 1\\
0 & 0 & 0 & 0 & 0 & 0 & 0 & 0 & 0 & 1 & 0 & 1\\
0 & 0 & 0 & 0 & 0 & 0 & 0 & 0 & 0 & 0 & 1 & 0\\
\end{array}\right)\)
}
\end{minipage}\hfill
\begin{minipage}[t]{0.32\textwidth}
\centering
\(M_{2}\)\par\smallskip
{\small\renewcommand{\arraystretch}{0.46}\setlength{\arraycolsep}{1.5pt}
\(\left(\begin{array}{ccc|ccc|ccc|ccc}
0 & 1 & 0 & 0 & 0 & 0 & 0 & 0 & 0 & 0 & 0 & 0\\
0 & 1 & 1 & 0 & 0 & 0 & 0 & 0 & 0 & 0 & 0 & 0\\
1 & 0 & 1 & 0 & 0 & 0 & 0 & 0 & 0 & 0 & 0 & 0\\\hline
0 & 0 & 0 & 0 & 1 & 0 & 0 & 0 & 0 & 0 & 0 & 0\\
0 & 0 & 0 & 0 & 1 & 1 & 0 & 0 & 0 & 0 & 0 & 0\\
0 & 0 & 0 & 1 & 0 & 1 & 0 & 0 & 0 & 0 & 0 & 0\\\hline
0 & 0 & 0 & 0 & 0 & 0 & 0 & 1 & 0 & 0 & 0 & 0\\
0 & 0 & 0 & 0 & 0 & 0 & 0 & 1 & 1 & 0 & 0 & 0\\
0 & 0 & 0 & 0 & 0 & 0 & 1 & 0 & 1 & 0 & 0 & 0\\\hline
0 & 0 & 0 & 0 & 0 & 0 & 0 & 0 & 0 & 0 & 1 & 0\\
0 & 0 & 0 & 0 & 0 & 0 & 0 & 0 & 0 & 0 & 1 & 1\\
0 & 0 & 0 & 0 & 0 & 0 & 0 & 0 & 0 & 1 & 0 & 1\\
\end{array}\right)\)
}
\end{minipage}\hfill
\begin{minipage}[t]{0.32\textwidth}
\centering
\(M_{3}\)\par\smallskip
{\small\renewcommand{\arraystretch}{0.46}\setlength{\arraycolsep}{1.5pt}
\(\left(\begin{array}{ccc|ccc|ccc|ccc}
1 & 0 & 1 & 0 & 0 & 0 & 0 & 0 & 0 & 0 & 0 & 0\\
1 & 1 & 1 & 0 & 0 & 0 & 0 & 0 & 0 & 0 & 0 & 0\\
0 & 1 & 1 & 0 & 0 & 0 & 0 & 0 & 0 & 0 & 0 & 0\\\hline
0 & 0 & 0 & 1 & 0 & 1 & 0 & 0 & 0 & 0 & 0 & 0\\
0 & 0 & 0 & 1 & 1 & 1 & 0 & 0 & 0 & 0 & 0 & 0\\
0 & 0 & 0 & 0 & 1 & 1 & 0 & 0 & 0 & 0 & 0 & 0\\\hline
0 & 0 & 0 & 0 & 0 & 0 & 1 & 0 & 1 & 0 & 0 & 0\\
0 & 0 & 0 & 0 & 0 & 0 & 1 & 1 & 1 & 0 & 0 & 0\\
0 & 0 & 0 & 0 & 0 & 0 & 0 & 1 & 1 & 0 & 0 & 0\\\hline
0 & 0 & 0 & 0 & 0 & 0 & 0 & 0 & 0 & 1 & 0 & 1\\
0 & 0 & 0 & 0 & 0 & 0 & 0 & 0 & 0 & 1 & 1 & 1\\
0 & 0 & 0 & 0 & 0 & 0 & 0 & 0 & 0 & 0 & 1 & 1\\
\end{array}\right)\)
}
\end{minipage}
\par\bigskip
\noindent
\begin{minipage}[t]{0.32\textwidth}
\centering
\(M_{4}\)\par\smallskip
{\small\renewcommand{\arraystretch}{0.46}\setlength{\arraycolsep}{1.5pt}
\(\left(\begin{array}{ccc|ccc|ccc|ccc}
0 & 0 & 0 & 0 & 0 & 1 & 1 & 1 & 1 & 1 & 1 & 0\\
0 & 0 & 0 & 1 & 0 & 1 & 1 & 0 & 0 & 0 & 0 & 1\\
0 & 0 & 0 & 0 & 1 & 0 & 1 & 1 & 0 & 1 & 0 & 0\\\hline
0 & 0 & 1 & 0 & 0 & 1 & 0 & 0 & 1 & 1 & 1 & 1\\
1 & 0 & 1 & 1 & 0 & 1 & 1 & 0 & 1 & 1 & 0 & 0\\
0 & 1 & 0 & 0 & 1 & 0 & 0 & 1 & 0 & 1 & 1 & 0\\\hline
0 & 0 & 0 & 0 & 0 & 0 & 0 & 0 & 0 & 0 & 0 & 1\\
0 & 0 & 0 & 0 & 0 & 0 & 0 & 0 & 0 & 1 & 0 & 1\\
0 & 0 & 0 & 0 & 0 & 0 & 0 & 0 & 0 & 0 & 1 & 0\\\hline
0 & 0 & 0 & 0 & 0 & 0 & 0 & 0 & 1 & 0 & 0 & 1\\
0 & 0 & 0 & 0 & 0 & 0 & 1 & 0 & 1 & 1 & 0 & 1\\
0 & 0 & 0 & 0 & 0 & 0 & 0 & 1 & 0 & 0 & 1 & 0\\
\end{array}\right)\)
}
\end{minipage}\hfill
\begin{minipage}[t]{0.32\textwidth}
\centering
\(M_{5}\)\par\smallskip
{\small\renewcommand{\arraystretch}{0.46}\setlength{\arraycolsep}{1.5pt}
\(\left(\begin{array}{ccc|ccc|ccc|ccc}
0 & 0 & 0 & 0 & 1 & 0 & 1 & 1 & 0 & 1 & 0 & 0\\
0 & 0 & 0 & 0 & 1 & 1 & 0 & 0 & 1 & 0 & 1 & 0\\
0 & 0 & 0 & 1 & 0 & 1 & 1 & 0 & 0 & 0 & 0 & 1\\\hline
0 & 1 & 0 & 0 & 1 & 0 & 0 & 1 & 0 & 1 & 1 & 0\\
0 & 1 & 1 & 0 & 1 & 1 & 0 & 1 & 1 & 0 & 0 & 1\\
1 & 0 & 1 & 1 & 0 & 1 & 1 & 0 & 1 & 1 & 0 & 0\\\hline
0 & 0 & 0 & 0 & 0 & 0 & 0 & 0 & 0 & 0 & 1 & 0\\
0 & 0 & 0 & 0 & 0 & 0 & 0 & 0 & 0 & 0 & 1 & 1\\
0 & 0 & 0 & 0 & 0 & 0 & 0 & 0 & 0 & 1 & 0 & 1\\\hline
0 & 0 & 0 & 0 & 0 & 0 & 0 & 1 & 0 & 0 & 1 & 0\\
0 & 0 & 0 & 0 & 0 & 0 & 0 & 1 & 1 & 0 & 1 & 1\\
0 & 0 & 0 & 0 & 0 & 0 & 1 & 0 & 1 & 1 & 0 & 1\\
\end{array}\right)\)
}
\end{minipage}\hfill
\begin{minipage}[t]{0.32\textwidth}
\centering
\(M_{6}\)\par\smallskip
{\small\renewcommand{\arraystretch}{0.46}\setlength{\arraycolsep}{1.5pt}
\(\left(\begin{array}{ccc|ccc|ccc|ccc}
0 & 0 & 0 & 1 & 0 & 1 & 1 & 0 & 0 & 0 & 0 & 1\\
0 & 0 & 0 & 1 & 1 & 1 & 0 & 1 & 0 & 1 & 0 & 1\\
0 & 0 & 0 & 0 & 1 & 1 & 0 & 0 & 1 & 0 & 1 & 0\\\hline
1 & 0 & 1 & 1 & 0 & 1 & 1 & 0 & 1 & 1 & 0 & 0\\
1 & 1 & 1 & 1 & 1 & 1 & 1 & 1 & 1 & 0 & 1 & 0\\
0 & 1 & 1 & 0 & 1 & 1 & 0 & 1 & 1 & 0 & 0 & 1\\\hline
0 & 0 & 0 & 0 & 0 & 0 & 0 & 0 & 0 & 1 & 0 & 1\\
0 & 0 & 0 & 0 & 0 & 0 & 0 & 0 & 0 & 1 & 1 & 1\\
0 & 0 & 0 & 0 & 0 & 0 & 0 & 0 & 0 & 0 & 1 & 1\\\hline
0 & 0 & 0 & 0 & 0 & 0 & 1 & 0 & 1 & 1 & 0 & 1\\
0 & 0 & 0 & 0 & 0 & 0 & 1 & 1 & 1 & 1 & 1 & 1\\
0 & 0 & 0 & 0 & 0 & 0 & 0 & 1 & 1 & 0 & 1 & 1\\
\end{array}\right)\)
}
\end{minipage}
\par\bigskip
\noindent
\begin{minipage}[t]{0.32\textwidth}
\centering
\(M_{7}\)\par\smallskip
{\small\renewcommand{\arraystretch}{0.46}\setlength{\arraycolsep}{1.5pt}
\(\left(\begin{array}{ccc|ccc|ccc|ccc}
0 & 0 & 1 & 0 & 0 & 1 & 0 & 1 & 1 & 0 & 0 & 1\\
0 & 1 & 0 & 0 & 1 & 0 & 1 & 1 & 1 & 0 & 1 & 0\\
1 & 1 & 1 & 1 & 1 & 1 & 0 & 0 & 1 & 1 & 1 & 1\\\hline
0 & 0 & 0 & 0 & 0 & 1 & 0 & 0 & 1 & 0 & 1 & 1\\
0 & 0 & 0 & 0 & 1 & 0 & 0 & 1 & 0 & 1 & 1 & 1\\
0 & 0 & 0 & 1 & 1 & 1 & 1 & 1 & 1 & 0 & 0 & 1\\\hline
0 & 0 & 0 & 0 & 0 & 0 & 0 & 0 & 1 & 0 & 0 & 1\\
0 & 0 & 0 & 0 & 0 & 0 & 0 & 1 & 0 & 0 & 1 & 0\\
0 & 0 & 0 & 0 & 0 & 0 & 1 & 1 & 1 & 1 & 1 & 1\\\hline
0 & 0 & 0 & 0 & 0 & 0 & 0 & 0 & 0 & 0 & 0 & 1\\
0 & 0 & 0 & 0 & 0 & 0 & 0 & 0 & 0 & 0 & 1 & 0\\
0 & 0 & 0 & 0 & 0 & 0 & 0 & 0 & 0 & 1 & 1 & 1\\
\end{array}\right)\)
}
\end{minipage}\hfill
\begin{minipage}[t]{0.32\textwidth}
\centering
\(M_{8}\)\par\smallskip
{\small\renewcommand{\arraystretch}{0.46}\setlength{\arraycolsep}{1.5pt}
\(\left(\begin{array}{ccc|ccc|ccc|ccc}
1 & 1 & 1 & 1 & 1 & 1 & 0 & 0 & 1 & 1 & 1 & 1\\
1 & 1 & 0 & 1 & 1 & 0 & 0 & 1 & 0 & 1 & 1 & 0\\
0 & 1 & 0 & 0 & 1 & 0 & 1 & 1 & 1 & 0 & 1 & 0\\\hline
0 & 0 & 0 & 1 & 1 & 1 & 1 & 1 & 1 & 0 & 0 & 1\\
0 & 0 & 0 & 1 & 1 & 0 & 1 & 1 & 0 & 0 & 1 & 0\\
0 & 0 & 0 & 0 & 1 & 0 & 0 & 1 & 0 & 1 & 1 & 1\\\hline
0 & 0 & 0 & 0 & 0 & 0 & 1 & 1 & 1 & 1 & 1 & 1\\
0 & 0 & 0 & 0 & 0 & 0 & 1 & 1 & 0 & 1 & 1 & 0\\
0 & 0 & 0 & 0 & 0 & 0 & 0 & 1 & 0 & 0 & 1 & 0\\\hline
0 & 0 & 0 & 0 & 0 & 0 & 0 & 0 & 0 & 1 & 1 & 1\\
0 & 0 & 0 & 0 & 0 & 0 & 0 & 0 & 0 & 1 & 1 & 0\\
0 & 0 & 0 & 0 & 0 & 0 & 0 & 0 & 0 & 0 & 1 & 0\\
\end{array}\right)\)
}
\end{minipage}\hfill
\begin{minipage}[t]{0.32\textwidth}
\centering
\(M_{9}\)\par\smallskip
{\small\renewcommand{\arraystretch}{0.46}\setlength{\arraycolsep}{1.5pt}
\(\left(\begin{array}{ccc|ccc|ccc|ccc}
0 & 1 & 0 & 0 & 1 & 0 & 1 & 1 & 1 & 0 & 1 & 0\\
1 & 0 & 1 & 1 & 0 & 1 & 1 & 1 & 0 & 1 & 0 & 1\\
1 & 1 & 0 & 1 & 1 & 0 & 0 & 1 & 0 & 1 & 1 & 0\\\hline
0 & 0 & 0 & 0 & 1 & 0 & 0 & 1 & 0 & 1 & 1 & 1\\
0 & 0 & 0 & 1 & 0 & 1 & 1 & 0 & 1 & 1 & 1 & 0\\
0 & 0 & 0 & 1 & 1 & 0 & 1 & 1 & 0 & 0 & 1 & 0\\\hline
0 & 0 & 0 & 0 & 0 & 0 & 0 & 1 & 0 & 0 & 1 & 0\\
0 & 0 & 0 & 0 & 0 & 0 & 1 & 0 & 1 & 1 & 0 & 1\\
0 & 0 & 0 & 0 & 0 & 0 & 1 & 1 & 0 & 1 & 1 & 0\\\hline
0 & 0 & 0 & 0 & 0 & 0 & 0 & 0 & 0 & 0 & 1 & 0\\
0 & 0 & 0 & 0 & 0 & 0 & 0 & 0 & 0 & 1 & 0 & 1\\
0 & 0 & 0 & 0 & 0 & 0 & 0 & 0 & 0 & 1 & 1 & 0\\
\end{array}\right)\)
}
\end{minipage}
\par\bigskip
\noindent
\begin{minipage}[t]{0.32\textwidth}
\centering
\(M_{10}\)\par\smallskip
{\small\renewcommand{\arraystretch}{0.46}\setlength{\arraycolsep}{1.5pt}
\(\left(\begin{array}{ccc|ccc|ccc|ccc}
0 & 0 & 0 & 0 & 0 & 1 & 1 & 0 & 0 & 0 & 1 & 0\\
0 & 0 & 0 & 0 & 1 & 0 & 0 & 0 & 1 & 1 & 0 & 1\\
0 & 0 & 0 & 1 & 1 & 1 & 0 & 1 & 1 & 1 & 1 & 0\\\hline
0 & 0 & 1 & 0 & 0 & 0 & 0 & 1 & 1 & 1 & 1 & 0\\
0 & 1 & 0 & 0 & 0 & 0 & 1 & 1 & 1 & 1 & 0 & 0\\
1 & 1 & 1 & 0 & 0 & 0 & 0 & 0 & 1 & 1 & 0 & 1\\\hline
0 & 0 & 0 & 0 & 0 & 0 & 0 & 0 & 0 & 0 & 0 & 1\\
0 & 0 & 0 & 0 & 0 & 0 & 0 & 0 & 0 & 0 & 1 & 0\\
0 & 0 & 0 & 0 & 0 & 0 & 0 & 0 & 0 & 1 & 1 & 1\\\hline
0 & 0 & 0 & 0 & 0 & 0 & 0 & 0 & 1 & 0 & 0 & 0\\
0 & 0 & 0 & 0 & 0 & 0 & 0 & 1 & 0 & 0 & 0 & 0\\
0 & 0 & 0 & 0 & 0 & 0 & 1 & 1 & 1 & 0 & 0 & 0\\
\end{array}\right)\)
}
\end{minipage}\hfill
\begin{minipage}[t]{0.32\textwidth}
\centering
\(M_{11}\)\par\smallskip
{\small\renewcommand{\arraystretch}{0.46}\setlength{\arraycolsep}{1.5pt}
\(\left(\begin{array}{ccc|ccc|ccc|ccc}
0 & 0 & 0 & 1 & 1 & 1 & 0 & 1 & 1 & 1 & 1 & 0\\
0 & 0 & 0 & 1 & 1 & 0 & 1 & 1 & 1 & 1 & 0 & 0\\
0 & 0 & 0 & 0 & 1 & 0 & 0 & 0 & 1 & 1 & 0 & 1\\\hline
1 & 1 & 1 & 0 & 0 & 0 & 0 & 0 & 1 & 1 & 0 & 1\\
1 & 1 & 0 & 0 & 0 & 0 & 0 & 1 & 0 & 0 & 1 & 1\\
0 & 1 & 0 & 0 & 0 & 0 & 1 & 1 & 1 & 1 & 0 & 0\\\hline
0 & 0 & 0 & 0 & 0 & 0 & 0 & 0 & 0 & 1 & 1 & 1\\
0 & 0 & 0 & 0 & 0 & 0 & 0 & 0 & 0 & 1 & 1 & 0\\
0 & 0 & 0 & 0 & 0 & 0 & 0 & 0 & 0 & 0 & 1 & 0\\\hline
0 & 0 & 0 & 0 & 0 & 0 & 1 & 1 & 1 & 0 & 0 & 0\\
0 & 0 & 0 & 0 & 0 & 0 & 1 & 1 & 0 & 0 & 0 & 0\\
0 & 0 & 0 & 0 & 0 & 0 & 0 & 1 & 0 & 0 & 0 & 0\\
\end{array}\right)\)
}
\end{minipage}\hfill
\begin{minipage}[t]{0.32\textwidth}
\centering
\(M_{12}\)\par\smallskip
{\small\renewcommand{\arraystretch}{0.46}\setlength{\arraycolsep}{1.5pt}
\(\left(\begin{array}{ccc|ccc|ccc|ccc}
0 & 0 & 0 & 0 & 1 & 0 & 0 & 0 & 1 & 1 & 0 & 1\\
0 & 0 & 0 & 1 & 0 & 1 & 0 & 1 & 0 & 0 & 1 & 1\\
0 & 0 & 0 & 1 & 1 & 0 & 1 & 1 & 1 & 1 & 0 & 0\\\hline
0 & 1 & 0 & 0 & 0 & 0 & 1 & 1 & 1 & 1 & 0 & 0\\
1 & 0 & 1 & 0 & 0 & 0 & 1 & 1 & 0 & 0 & 0 & 1\\
1 & 1 & 0 & 0 & 0 & 0 & 0 & 1 & 0 & 0 & 1 & 1\\\hline
0 & 0 & 0 & 0 & 0 & 0 & 0 & 0 & 0 & 0 & 1 & 0\\
0 & 0 & 0 & 0 & 0 & 0 & 0 & 0 & 0 & 1 & 0 & 1\\
0 & 0 & 0 & 0 & 0 & 0 & 0 & 0 & 0 & 1 & 1 & 0\\\hline
0 & 0 & 0 & 0 & 0 & 0 & 0 & 1 & 0 & 0 & 0 & 0\\
0 & 0 & 0 & 0 & 0 & 0 & 1 & 0 & 1 & 0 & 0 & 0\\
0 & 0 & 0 & 0 & 0 & 0 & 1 & 1 & 0 & 0 & 0 & 0\\
\end{array}\right)\)
}
\end{minipage}
\par\bigskip
\noindent
\begin{minipage}[t]{0.32\textwidth}
\centering
\(M_{13}\)\par\smallskip
{\small\renewcommand{\arraystretch}{0.46}\setlength{\arraycolsep}{1.5pt}
\(\left(\begin{array}{ccc|ccc|ccc|ccc}
0 & 0 & 1 & 0 & 0 & 0 & 1 & 0 & 0 & 1 & 0 & 1\\
1 & 1 & 1 & 0 & 0 & 0 & 0 & 1 & 1 & 1 & 0 & 0\\
1 & 0 & 1 & 0 & 0 & 0 & 0 & 1 & 0 & 1 & 1 & 1\\\hline
0 & 0 & 0 & 0 & 0 & 1 & 0 & 0 & 0 & 1 & 0 & 0\\
0 & 0 & 0 & 1 & 1 & 1 & 0 & 0 & 0 & 0 & 1 & 1\\
0 & 0 & 0 & 1 & 0 & 1 & 0 & 0 & 0 & 0 & 1 & 0\\\hline
0 & 0 & 0 & 0 & 0 & 0 & 0 & 0 & 1 & 0 & 0 & 0\\
0 & 0 & 0 & 0 & 0 & 0 & 1 & 1 & 1 & 0 & 0 & 0\\
0 & 0 & 0 & 0 & 0 & 0 & 1 & 0 & 1 & 0 & 0 & 0\\\hline
0 & 0 & 0 & 0 & 0 & 0 & 0 & 0 & 0 & 0 & 0 & 1\\
0 & 0 & 0 & 0 & 0 & 0 & 0 & 0 & 0 & 1 & 1 & 1\\
0 & 0 & 0 & 0 & 0 & 0 & 0 & 0 & 0 & 1 & 0 & 1\\
\end{array}\right)\)
}
\end{minipage}\hfill
\begin{minipage}[t]{0.32\textwidth}
\centering
\(M_{14}\)\par\smallskip
{\small\renewcommand{\arraystretch}{0.46}\setlength{\arraycolsep}{1.5pt}
\(\left(\begin{array}{ccc|ccc|ccc|ccc}
1 & 0 & 1 & 0 & 0 & 0 & 0 & 1 & 0 & 1 & 1 & 1\\
1 & 0 & 0 & 0 & 0 & 0 & 1 & 1 & 0 & 0 & 1 & 0\\
1 & 1 & 1 & 0 & 0 & 0 & 0 & 1 & 1 & 1 & 0 & 0\\\hline
0 & 0 & 0 & 1 & 0 & 1 & 0 & 0 & 0 & 0 & 1 & 0\\
0 & 0 & 0 & 1 & 0 & 0 & 0 & 0 & 0 & 1 & 1 & 0\\
0 & 0 & 0 & 1 & 1 & 1 & 0 & 0 & 0 & 0 & 1 & 1\\\hline
0 & 0 & 0 & 0 & 0 & 0 & 1 & 0 & 1 & 0 & 0 & 0\\
0 & 0 & 0 & 0 & 0 & 0 & 1 & 0 & 0 & 0 & 0 & 0\\
0 & 0 & 0 & 0 & 0 & 0 & 1 & 1 & 1 & 0 & 0 & 0\\\hline
0 & 0 & 0 & 0 & 0 & 0 & 0 & 0 & 0 & 1 & 0 & 1\\
0 & 0 & 0 & 0 & 0 & 0 & 0 & 0 & 0 & 1 & 0 & 0\\
0 & 0 & 0 & 0 & 0 & 0 & 0 & 0 & 0 & 1 & 1 & 1\\
\end{array}\right)\)
}
\end{minipage}\hfill
\begin{minipage}[t]{0.32\textwidth}
\centering
\(M_{15}\)\par\smallskip
{\small\renewcommand{\arraystretch}{0.46}\setlength{\arraycolsep}{1.5pt}
\(\left(\begin{array}{ccc|ccc|ccc|ccc}
1 & 1 & 1 & 0 & 0 & 0 & 0 & 1 & 1 & 1 & 0 & 0\\
0 & 1 & 0 & 0 & 0 & 0 & 0 & 0 & 1 & 0 & 1 & 1\\
1 & 0 & 0 & 0 & 0 & 0 & 1 & 1 & 0 & 0 & 1 & 0\\\hline
0 & 0 & 0 & 1 & 1 & 1 & 0 & 0 & 0 & 0 & 1 & 1\\
0 & 0 & 0 & 0 & 1 & 0 & 0 & 0 & 0 & 0 & 0 & 1\\
0 & 0 & 0 & 1 & 0 & 0 & 0 & 0 & 0 & 1 & 1 & 0\\\hline
0 & 0 & 0 & 0 & 0 & 0 & 1 & 1 & 1 & 0 & 0 & 0\\
0 & 0 & 0 & 0 & 0 & 0 & 0 & 1 & 0 & 0 & 0 & 0\\
0 & 0 & 0 & 0 & 0 & 0 & 1 & 0 & 0 & 0 & 0 & 0\\\hline
0 & 0 & 0 & 0 & 0 & 0 & 0 & 0 & 0 & 1 & 1 & 1\\
0 & 0 & 0 & 0 & 0 & 0 & 0 & 0 & 0 & 0 & 1 & 0\\
0 & 0 & 0 & 0 & 0 & 0 & 0 & 0 & 0 & 1 & 0 & 0\\
\end{array}\right)\)
}
\end{minipage}
\par\bigskip
\noindent
\begin{minipage}[t]{0.32\textwidth}
\centering
\(M_{16}\)\par\smallskip
{\small\renewcommand{\arraystretch}{0.46}\setlength{\arraycolsep}{1.5pt}
\(\left(\begin{array}{ccc|ccc|ccc|ccc}
0 & 0 & 0 & 0 & 0 & 1 & 0 & 1 & 0 & 1 & 1 & 1\\
0 & 0 & 0 & 1 & 1 & 1 & 1 & 1 & 0 & 0 & 1 & 0\\
0 & 0 & 0 & 1 & 0 & 1 & 0 & 1 & 1 & 1 & 0 & 0\\\hline
0 & 0 & 1 & 0 & 0 & 1 & 1 & 0 & 1 & 0 & 1 & 1\\
1 & 1 & 1 & 1 & 1 & 1 & 1 & 0 & 0 & 0 & 0 & 1\\
1 & 0 & 1 & 1 & 0 & 1 & 1 & 1 & 1 & 1 & 1 & 0\\\hline
0 & 0 & 0 & 0 & 0 & 0 & 0 & 0 & 0 & 0 & 0 & 1\\
0 & 0 & 0 & 0 & 0 & 0 & 0 & 0 & 0 & 1 & 1 & 1\\
0 & 0 & 0 & 0 & 0 & 0 & 0 & 0 & 0 & 1 & 0 & 1\\\hline
0 & 0 & 0 & 0 & 0 & 0 & 0 & 0 & 1 & 0 & 0 & 1\\
0 & 0 & 0 & 0 & 0 & 0 & 1 & 1 & 1 & 1 & 1 & 1\\
0 & 0 & 0 & 0 & 0 & 0 & 1 & 0 & 1 & 1 & 0 & 1\\
\end{array}\right)\)
}
\end{minipage}\hfill
\begin{minipage}[t]{0.32\textwidth}
\centering
\(M_{17}\)\par\smallskip
{\small\renewcommand{\arraystretch}{0.46}\setlength{\arraycolsep}{1.5pt}
\(\left(\begin{array}{ccc|ccc|ccc|ccc}
0 & 0 & 0 & 1 & 0 & 1 & 0 & 1 & 1 & 1 & 0 & 0\\
0 & 0 & 0 & 1 & 0 & 0 & 0 & 0 & 1 & 0 & 1 & 1\\
0 & 0 & 0 & 1 & 1 & 1 & 1 & 1 & 0 & 0 & 1 & 0\\\hline
1 & 0 & 1 & 1 & 0 & 1 & 1 & 1 & 1 & 1 & 1 & 0\\
1 & 0 & 0 & 1 & 0 & 0 & 0 & 1 & 0 & 1 & 0 & 1\\
1 & 1 & 1 & 1 & 1 & 1 & 1 & 0 & 0 & 0 & 0 & 1\\\hline
0 & 0 & 0 & 0 & 0 & 0 & 0 & 0 & 0 & 1 & 0 & 1\\
0 & 0 & 0 & 0 & 0 & 0 & 0 & 0 & 0 & 1 & 0 & 0\\
0 & 0 & 0 & 0 & 0 & 0 & 0 & 0 & 0 & 1 & 1 & 1\\\hline
0 & 0 & 0 & 0 & 0 & 0 & 1 & 0 & 1 & 1 & 0 & 1\\
0 & 0 & 0 & 0 & 0 & 0 & 1 & 0 & 0 & 1 & 0 & 0\\
0 & 0 & 0 & 0 & 0 & 0 & 1 & 1 & 1 & 1 & 1 & 1\\
\end{array}\right)\)
}
\end{minipage}\hfill
\begin{minipage}[t]{0.32\textwidth}
\centering
\(M_{18}\)\par\smallskip
{\small\renewcommand{\arraystretch}{0.46}\setlength{\arraycolsep}{1.5pt}
\(\left(\begin{array}{ccc|ccc|ccc|ccc}
0 & 0 & 0 & 1 & 1 & 1 & 1 & 1 & 0 & 0 & 1 & 0\\
0 & 0 & 0 & 0 & 1 & 0 & 1 & 0 & 1 & 1 & 1 & 0\\
0 & 0 & 0 & 1 & 0 & 0 & 0 & 0 & 1 & 0 & 1 & 1\\\hline
1 & 1 & 1 & 1 & 1 & 1 & 1 & 0 & 0 & 0 & 0 & 1\\
0 & 1 & 0 & 0 & 1 & 0 & 0 & 1 & 1 & 1 & 1 & 1\\
1 & 0 & 0 & 1 & 0 & 0 & 0 & 1 & 0 & 1 & 0 & 1\\\hline
0 & 0 & 0 & 0 & 0 & 0 & 0 & 0 & 0 & 1 & 1 & 1\\
0 & 0 & 0 & 0 & 0 & 0 & 0 & 0 & 0 & 0 & 1 & 0\\
0 & 0 & 0 & 0 & 0 & 0 & 0 & 0 & 0 & 1 & 0 & 0\\\hline
0 & 0 & 0 & 0 & 0 & 0 & 1 & 1 & 1 & 1 & 1 & 1\\
0 & 0 & 0 & 0 & 0 & 0 & 0 & 1 & 0 & 0 & 1 & 0\\
0 & 0 & 0 & 0 & 0 & 0 & 1 & 0 & 0 & 1 & 0 & 0\\
\end{array}\right)\)
}
\end{minipage}
\par\bigskip
\noindent
\begin{minipage}[t]{0.32\textwidth}
\centering
\(M_{19}\)\par\smallskip
{\small\renewcommand{\arraystretch}{0.46}\setlength{\arraycolsep}{1.5pt}
\(\left(\begin{array}{ccc|ccc|ccc|ccc}
0 & 0 & 0 & 0 & 0 & 1 & 0 & 0 & 0 & 0 & 0 & 0\\
0 & 0 & 0 & 1 & 0 & 1 & 0 & 0 & 0 & 0 & 0 & 0\\
0 & 0 & 0 & 0 & 1 & 0 & 0 & 0 & 0 & 0 & 0 & 0\\\hline
0 & 0 & 0 & 0 & 0 & 0 & 0 & 0 & 1 & 0 & 0 & 0\\
0 & 0 & 0 & 0 & 0 & 0 & 1 & 0 & 1 & 0 & 0 & 0\\
0 & 0 & 0 & 0 & 0 & 0 & 0 & 1 & 0 & 0 & 0 & 0\\\hline
0 & 0 & 0 & 0 & 0 & 0 & 0 & 0 & 0 & 0 & 0 & 1\\
0 & 0 & 0 & 0 & 0 & 0 & 0 & 0 & 0 & 1 & 0 & 1\\
0 & 0 & 0 & 0 & 0 & 0 & 0 & 0 & 0 & 0 & 1 & 0\\\hline
0 & 0 & 0 & 0 & 0 & 0 & 0 & 0 & 0 & 0 & 0 & 0\\
0 & 0 & 0 & 0 & 0 & 0 & 0 & 0 & 0 & 0 & 0 & 0\\
0 & 0 & 0 & 0 & 0 & 0 & 0 & 0 & 0 & 0 & 0 & 0\\
\end{array}\right)\)
}
\end{minipage}\hfill
\begin{minipage}[t]{0.32\textwidth}
\centering
\(M_{20}\)\par\smallskip
{\small\renewcommand{\arraystretch}{0.46}\setlength{\arraycolsep}{1.5pt}
\(\left(\begin{array}{ccc|ccc|ccc|ccc}
0 & 0 & 0 & 0 & 1 & 0 & 0 & 0 & 0 & 0 & 0 & 0\\
0 & 0 & 0 & 0 & 1 & 1 & 0 & 0 & 0 & 0 & 0 & 0\\
0 & 0 & 0 & 1 & 0 & 1 & 0 & 0 & 0 & 0 & 0 & 0\\\hline
0 & 0 & 0 & 0 & 0 & 0 & 0 & 1 & 0 & 0 & 0 & 0\\
0 & 0 & 0 & 0 & 0 & 0 & 0 & 1 & 1 & 0 & 0 & 0\\
0 & 0 & 0 & 0 & 0 & 0 & 1 & 0 & 1 & 0 & 0 & 0\\\hline
0 & 0 & 0 & 0 & 0 & 0 & 0 & 0 & 0 & 0 & 1 & 0\\
0 & 0 & 0 & 0 & 0 & 0 & 0 & 0 & 0 & 0 & 1 & 1\\
0 & 0 & 0 & 0 & 0 & 0 & 0 & 0 & 0 & 1 & 0 & 1\\\hline
0 & 0 & 0 & 0 & 0 & 0 & 0 & 0 & 0 & 0 & 0 & 0\\
0 & 0 & 0 & 0 & 0 & 0 & 0 & 0 & 0 & 0 & 0 & 0\\
0 & 0 & 0 & 0 & 0 & 0 & 0 & 0 & 0 & 0 & 0 & 0\\
\end{array}\right)\)
}
\end{minipage}\hfill
\begin{minipage}[t]{0.32\textwidth}
\centering
\(M_{21}\)\par\smallskip
{\small\renewcommand{\arraystretch}{0.46}\setlength{\arraycolsep}{1.5pt}
\(\left(\begin{array}{ccc|ccc|ccc|ccc}
0 & 0 & 0 & 1 & 0 & 1 & 0 & 0 & 0 & 0 & 0 & 0\\
0 & 0 & 0 & 1 & 1 & 1 & 0 & 0 & 0 & 0 & 0 & 0\\
0 & 0 & 0 & 0 & 1 & 1 & 0 & 0 & 0 & 0 & 0 & 0\\\hline
0 & 0 & 0 & 0 & 0 & 0 & 1 & 0 & 1 & 0 & 0 & 0\\
0 & 0 & 0 & 0 & 0 & 0 & 1 & 1 & 1 & 0 & 0 & 0\\
0 & 0 & 0 & 0 & 0 & 0 & 0 & 1 & 1 & 0 & 0 & 0\\\hline
0 & 0 & 0 & 0 & 0 & 0 & 0 & 0 & 0 & 1 & 0 & 1\\
0 & 0 & 0 & 0 & 0 & 0 & 0 & 0 & 0 & 1 & 1 & 1\\
0 & 0 & 0 & 0 & 0 & 0 & 0 & 0 & 0 & 0 & 1 & 1\\\hline
0 & 0 & 0 & 0 & 0 & 0 & 0 & 0 & 0 & 0 & 0 & 0\\
0 & 0 & 0 & 0 & 0 & 0 & 0 & 0 & 0 & 0 & 0 & 0\\
0 & 0 & 0 & 0 & 0 & 0 & 0 & 0 & 0 & 0 & 0 & 0\\
\end{array}\right)\)
}
\end{minipage}
\par\bigskip
\noindent
\begin{minipage}[t]{0.32\textwidth}
\centering
\(M_{22}\)\par\smallskip
{\small\renewcommand{\arraystretch}{0.46}\setlength{\arraycolsep}{1.5pt}
\(\left(\begin{array}{ccc|ccc|ccc|ccc}
0 & 0 & 0 & 0 & 0 & 0 & 0 & 0 & 1 & 1 & 1 & 1\\
0 & 0 & 0 & 0 & 0 & 0 & 1 & 0 & 1 & 1 & 0 & 0\\
0 & 0 & 0 & 0 & 0 & 0 & 0 & 1 & 0 & 1 & 1 & 0\\\hline
0 & 0 & 0 & 0 & 0 & 1 & 0 & 0 & 1 & 0 & 0 & 1\\
0 & 0 & 0 & 1 & 0 & 1 & 1 & 0 & 1 & 1 & 0 & 1\\
0 & 0 & 0 & 0 & 1 & 0 & 0 & 1 & 0 & 0 & 1 & 0\\\hline
0 & 0 & 0 & 0 & 0 & 0 & 0 & 0 & 0 & 0 & 0 & 0\\
0 & 0 & 0 & 0 & 0 & 0 & 0 & 0 & 0 & 0 & 0 & 0\\
0 & 0 & 0 & 0 & 0 & 0 & 0 & 0 & 0 & 0 & 0 & 0\\\hline
0 & 0 & 0 & 0 & 0 & 0 & 0 & 0 & 0 & 0 & 0 & 1\\
0 & 0 & 0 & 0 & 0 & 0 & 0 & 0 & 0 & 1 & 0 & 1\\
0 & 0 & 0 & 0 & 0 & 0 & 0 & 0 & 0 & 0 & 1 & 0\\
\end{array}\right)\)
}
\end{minipage}\hfill
\begin{minipage}[t]{0.32\textwidth}
\centering
\(M_{23}\)\par\smallskip
{\small\renewcommand{\arraystretch}{0.46}\setlength{\arraycolsep}{1.5pt}
\(\left(\begin{array}{ccc|ccc|ccc|ccc}
0 & 0 & 0 & 0 & 0 & 0 & 0 & 1 & 0 & 1 & 1 & 0\\
0 & 0 & 0 & 0 & 0 & 0 & 0 & 1 & 1 & 0 & 0 & 1\\
0 & 0 & 0 & 0 & 0 & 0 & 1 & 0 & 1 & 1 & 0 & 0\\\hline
0 & 0 & 0 & 0 & 1 & 0 & 0 & 1 & 0 & 0 & 1 & 0\\
0 & 0 & 0 & 0 & 1 & 1 & 0 & 1 & 1 & 0 & 1 & 1\\
0 & 0 & 0 & 1 & 0 & 1 & 1 & 0 & 1 & 1 & 0 & 1\\\hline
0 & 0 & 0 & 0 & 0 & 0 & 0 & 0 & 0 & 0 & 0 & 0\\
0 & 0 & 0 & 0 & 0 & 0 & 0 & 0 & 0 & 0 & 0 & 0\\
0 & 0 & 0 & 0 & 0 & 0 & 0 & 0 & 0 & 0 & 0 & 0\\\hline
0 & 0 & 0 & 0 & 0 & 0 & 0 & 0 & 0 & 0 & 1 & 0\\
0 & 0 & 0 & 0 & 0 & 0 & 0 & 0 & 0 & 0 & 1 & 1\\
0 & 0 & 0 & 0 & 0 & 0 & 0 & 0 & 0 & 1 & 0 & 1\\
\end{array}\right)\)
}
\end{minipage}\hfill
\begin{minipage}[t]{0.32\textwidth}
\centering
\(M_{24}\)\par\smallskip
{\small\renewcommand{\arraystretch}{0.46}\setlength{\arraycolsep}{1.5pt}
\(\left(\begin{array}{ccc|ccc|ccc|ccc}
0 & 0 & 0 & 0 & 0 & 0 & 1 & 0 & 1 & 1 & 0 & 0\\
0 & 0 & 0 & 0 & 0 & 0 & 1 & 1 & 1 & 0 & 1 & 0\\
0 & 0 & 0 & 0 & 0 & 0 & 0 & 1 & 1 & 0 & 0 & 1\\\hline
0 & 0 & 0 & 1 & 0 & 1 & 1 & 0 & 1 & 1 & 0 & 1\\
0 & 0 & 0 & 1 & 1 & 1 & 1 & 1 & 1 & 1 & 1 & 1\\
0 & 0 & 0 & 0 & 1 & 1 & 0 & 1 & 1 & 0 & 1 & 1\\\hline
0 & 0 & 0 & 0 & 0 & 0 & 0 & 0 & 0 & 0 & 0 & 0\\
0 & 0 & 0 & 0 & 0 & 0 & 0 & 0 & 0 & 0 & 0 & 0\\
0 & 0 & 0 & 0 & 0 & 0 & 0 & 0 & 0 & 0 & 0 & 0\\\hline
0 & 0 & 0 & 0 & 0 & 0 & 0 & 0 & 0 & 1 & 0 & 1\\
0 & 0 & 0 & 0 & 0 & 0 & 0 & 0 & 0 & 1 & 1 & 1\\
0 & 0 & 0 & 0 & 0 & 0 & 0 & 0 & 0 & 0 & 1 & 1\\
\end{array}\right)\)
}
\end{minipage}
\par\bigskip
\noindent
\begin{minipage}[t]{0.32\textwidth}
\centering
\(M_{25}\)\par\smallskip
{\small\renewcommand{\arraystretch}{0.46}\setlength{\arraycolsep}{1.5pt}
\(\left(\begin{array}{ccc|ccc|ccc|ccc}
0 & 0 & 0 & 0 & 0 & 0 & 1 & 1 & 1 & 1 & 0 & 0\\
0 & 0 & 0 & 0 & 0 & 0 & 1 & 0 & 0 & 0 & 1 & 0\\
0 & 0 & 0 & 0 & 0 & 0 & 1 & 1 & 0 & 0 & 0 & 1\\\hline
0 & 0 & 0 & 0 & 0 & 0 & 0 & 0 & 1 & 0 & 0 & 1\\
0 & 0 & 0 & 0 & 0 & 0 & 1 & 0 & 1 & 1 & 0 & 1\\
0 & 0 & 0 & 0 & 0 & 0 & 0 & 1 & 0 & 0 & 1 & 0\\\hline
0 & 0 & 0 & 0 & 0 & 1 & 0 & 0 & 0 & 0 & 0 & 1\\
0 & 0 & 0 & 1 & 0 & 1 & 0 & 0 & 0 & 1 & 0 & 1\\
0 & 0 & 0 & 0 & 1 & 0 & 0 & 0 & 0 & 0 & 1 & 0\\\hline
0 & 0 & 0 & 0 & 0 & 0 & 0 & 0 & 1 & 0 & 0 & 0\\
0 & 0 & 0 & 0 & 0 & 0 & 1 & 0 & 1 & 0 & 0 & 0\\
0 & 0 & 0 & 0 & 0 & 0 & 0 & 1 & 0 & 0 & 0 & 0\\
\end{array}\right)\)
}
\end{minipage}\hfill
\begin{minipage}[t]{0.32\textwidth}
\centering
\(M_{26}\)\par\smallskip
{\small\renewcommand{\arraystretch}{0.46}\setlength{\arraycolsep}{1.5pt}
\(\left(\begin{array}{ccc|ccc|ccc|ccc}
0 & 0 & 0 & 0 & 0 & 0 & 1 & 1 & 0 & 0 & 0 & 1\\
0 & 0 & 0 & 0 & 0 & 0 & 0 & 0 & 1 & 1 & 0 & 1\\
0 & 0 & 0 & 0 & 0 & 0 & 1 & 0 & 0 & 0 & 1 & 0\\\hline
0 & 0 & 0 & 0 & 0 & 0 & 0 & 1 & 0 & 0 & 1 & 0\\
0 & 0 & 0 & 0 & 0 & 0 & 0 & 1 & 1 & 0 & 1 & 1\\
0 & 0 & 0 & 0 & 0 & 0 & 1 & 0 & 1 & 1 & 0 & 1\\\hline
0 & 0 & 0 & 0 & 1 & 0 & 0 & 0 & 0 & 0 & 1 & 0\\
0 & 0 & 0 & 0 & 1 & 1 & 0 & 0 & 0 & 0 & 1 & 1\\
0 & 0 & 0 & 1 & 0 & 1 & 0 & 0 & 0 & 1 & 0 & 1\\\hline
0 & 0 & 0 & 0 & 0 & 0 & 0 & 1 & 0 & 0 & 0 & 0\\
0 & 0 & 0 & 0 & 0 & 0 & 0 & 1 & 1 & 0 & 0 & 0\\
0 & 0 & 0 & 0 & 0 & 0 & 1 & 0 & 1 & 0 & 0 & 0\\
\end{array}\right)\)
}
\end{minipage}\hfill
\begin{minipage}[t]{0.32\textwidth}
\centering
\(M_{27}\)\par\smallskip
{\small\renewcommand{\arraystretch}{0.46}\setlength{\arraycolsep}{1.5pt}
\(\left(\begin{array}{ccc|ccc|ccc|ccc}
0 & 0 & 0 & 0 & 0 & 0 & 1 & 0 & 0 & 0 & 1 & 0\\
0 & 0 & 0 & 0 & 0 & 0 & 0 & 1 & 0 & 0 & 1 & 1\\
0 & 0 & 0 & 0 & 0 & 0 & 0 & 0 & 1 & 1 & 0 & 1\\\hline
0 & 0 & 0 & 0 & 0 & 0 & 1 & 0 & 1 & 1 & 0 & 1\\
0 & 0 & 0 & 0 & 0 & 0 & 1 & 1 & 1 & 1 & 1 & 1\\
0 & 0 & 0 & 0 & 0 & 0 & 0 & 1 & 1 & 0 & 1 & 1\\\hline
0 & 0 & 0 & 1 & 0 & 1 & 0 & 0 & 0 & 1 & 0 & 1\\
0 & 0 & 0 & 1 & 1 & 1 & 0 & 0 & 0 & 1 & 1 & 1\\
0 & 0 & 0 & 0 & 1 & 1 & 0 & 0 & 0 & 0 & 1 & 1\\\hline
0 & 0 & 0 & 0 & 0 & 0 & 1 & 0 & 1 & 0 & 0 & 0\\
0 & 0 & 0 & 0 & 0 & 0 & 1 & 1 & 1 & 0 & 0 & 0\\
0 & 0 & 0 & 0 & 0 & 0 & 0 & 1 & 1 & 0 & 0 & 0\\
\end{array}\right)\)
}
\end{minipage}
\par\bigskip

\section{Basis matrices for Example~\ref{ex3_2}}
\noindent Matrices \(M_1, M_2, M_3\):
\[
\scalebox{0.95}{$
	{\renewcommand{\arraystretch}{0.46}\setlength{\arraycolsep}{1.5pt}
	\left(\begin{array}{ccc|ccc|ccc|ccc}
		0 & 1 & 0 & 0 & 0 & 0 & 0 & 0 & 0 & 0 & 0 & 0 \\
		0 & 1 & 1 & 0 & 0 & 0 & 0 & 0 & 0 & 0 & 0 & 0 \\
		1 & 0 & 1 & 0 & 0 & 0 & 0 & 0 & 0 & 0 & 0 & 0 \\\hline
		0 & 0 & 0 & 0 & 1 & 0 & 0 & 0 & 0 & 0 & 0 & 0 \\
		0 & 0 & 0 & 0 & 1 & 1 & 0 & 0 & 0 & 0 & 0 & 0 \\
		0 & 0 & 0 & 1 & 0 & 1 & 0 & 0 & 0 & 0 & 0 & 0 \\\hline
		0 & 0 & 0 & 0 & 0 & 0 & 0 & 1 & 0 & 0 & 0 & 0 \\
		0 & 0 & 0 & 0 & 0 & 0 & 0 & 1 & 1 & 0 & 0 & 0 \\
		0 & 0 & 0 & 0 & 0 & 0 & 1 & 0 & 1 & 0 & 0 & 0 \\\hline
		0 & 0 & 0 & 0 & 0 & 0 & 0 & 0 & 0 & 0 & 1 & 0 \\
		0 & 0 & 0 & 0 & 0 & 0 & 0 & 0 & 0 & 0 & 1 & 1 \\
		0 & 0 & 0 & 0 & 0 & 0 & 0 & 0 & 0 & 1 & 0 & 1 \\
	\end{array}\right)
	\quad\quad
    \left(\begin{array}{ccc|ccc|ccc|ccc}
		1 & 0 & 1 & 0 & 0 & 0 & 0 & 0 & 0 & 0 & 0 & 0 \\
		1 & 1 & 1 & 0 & 0 & 0 & 0 & 0 & 0 & 0 & 0 & 0 \\
		0 & 1 & 1 & 0 & 0 & 0 & 0 & 0 & 0 & 0 & 0 & 0 \\\hline
		0 & 0 & 0 & 1 & 0 & 1 & 0 & 0 & 0 & 0 & 0 & 0 \\
		0 & 0 & 0 & 1 & 1 & 1 & 0 & 0 & 0 & 0 & 0 & 0 \\
		0 & 0 & 0 & 0 & 1 & 1 & 0 & 0 & 0 & 0 & 0 & 0 \\\hline
		0 & 0 & 0 & 0 & 0 & 0 & 1 & 0 & 1 & 0 & 0 & 0 \\
		0 & 0 & 0 & 0 & 0 & 0 & 1 & 1 & 1 & 0 & 0 & 0 \\
		0 & 0 & 0 & 0 & 0 & 0 & 0 & 1 & 1 & 0 & 0 & 0 \\\hline
		0 & 0 & 0 & 0 & 0 & 0 & 0 & 0 & 0 & 1 & 0 & 1 \\
		0 & 0 & 0 & 0 & 0 & 0 & 0 & 0 & 0 & 1 & 1 & 1 \\
		0 & 0 & 0 & 0 & 0 & 0 & 0 & 0 & 0 & 0 & 1 & 1 \\
	\end{array}\right)
	\quad\quad
	\left(\begin{array}{ccc|ccc|ccc|ccc}
		0 & 1 & 1 & 0 & 0 & 0 & 0 & 0 & 0 & 0 & 0 & 0 \\
		1 & 1 & 0 & 0 & 0 & 0 & 0 & 0 & 0 & 0 & 0 & 0 \\
		1 & 1 & 1 & 0 & 0 & 0 & 0 & 0 & 0 & 0 & 0 & 0 \\\hline
		0 & 0 & 0 & 0 & 1 & 1 & 0 & 0 & 0 & 0 & 0 & 0 \\
		0 & 0 & 0 & 1 & 1 & 0 & 0 & 0 & 0 & 0 & 0 & 0 \\
		0 & 0 & 0 & 1 & 1 & 1 & 0 & 0 & 0 & 0 & 0 & 0 \\\hline
		0 & 0 & 0 & 0 & 0 & 0 & 0 & 1 & 1 & 0 & 0 & 0 \\
		0 & 0 & 0 & 0 & 0 & 0 & 1 & 1 & 0 & 0 & 0 & 0 \\
		0 & 0 & 0 & 0 & 0 & 0 & 1 & 1 & 1 & 0 & 0 & 0 \\\hline
		0 & 0 & 0 & 0 & 0 & 0 & 0 & 0 & 0 & 0 & 1 & 1 \\
		0 & 0 & 0 & 0 & 0 & 0 & 0 & 0 & 0 & 1 & 1 & 0 \\
		0 & 0 & 0 & 0 & 0 & 0 & 0 & 0 & 0 & 1 & 1 & 1 \\
	\end{array}\right)
    }
	$}
\]
\vspace{1em}
\noindent Matrices \(M_4, M_5, M_6\):
\[
\scalebox{0.95}{$
	{\renewcommand{\arraystretch}{0.46}\setlength{\arraycolsep}{1.5pt}
	\left(\begin{array}{ccc|ccc|ccc|ccc}
		0 & 1 & 0 & 0 & 1 & 0 & 0 & 0 & 1 & 0 & 1 & 0 \\
		1 & 0 & 1 & 1 & 0 & 1 & 0 & 1 & 0 & 1 & 0 & 1 \\
		1 & 1 & 0 & 1 & 1 & 0 & 1 & 1 & 1 & 1 & 1 & 0 \\\hline
		0 & 0 & 0 & 0 & 1 & 0 & 0 & 1 & 0 & 0 & 0 & 1 \\
		0 & 0 & 0 & 1 & 0 & 1 & 1 & 0 & 1 & 0 & 1 & 0 \\
		0 & 0 & 0 & 1 & 1 & 0 & 1 & 1 & 0 & 1 & 1 & 1 \\\hline
		0 & 0 & 0 & 0 & 0 & 0 & 0 & 1 & 0 & 0 & 1 & 0 \\
		0 & 0 & 0 & 0 & 0 & 0 & 1 & 0 & 1 & 1 & 0 & 1 \\
		0 & 0 & 0 & 0 & 0 & 0 & 1 & 1 & 0 & 1 & 1 & 0 \\\hline
		0 & 0 & 0 & 0 & 0 & 0 & 0 & 0 & 0 & 0 & 1 & 0 \\
		0 & 0 & 0 & 0 & 0 & 0 & 0 & 0 & 0 & 1 & 0 & 1 \\
		0 & 0 & 0 & 0 & 0 & 0 & 0 & 0 & 0 & 1 & 1 & 0 \\
	\end{array}\right)
	\quad\quad
	\left(\begin{array}{ccc|ccc|ccc|ccc}
		1 & 1 & 0 & 1 & 1 & 0 & 1 & 1 & 1 & 1 & 1 & 0 \\
		1 & 0 & 0 & 1 & 0 & 0 & 1 & 1 & 0 & 1 & 0 & 0 \\
		1 & 0 & 1 & 1 & 0 & 1 & 0 & 1 & 0 & 1 & 0 & 1 \\\hline
		0 & 0 & 0 & 1 & 1 & 0 & 1 & 1 & 0 & 1 & 1 & 1 \\
		0 & 0 & 0 & 1 & 0 & 0 & 1 & 0 & 0 & 1 & 1 & 0 \\
		0 & 0 & 0 & 1 & 0 & 1 & 1 & 0 & 1 & 0 & 1 & 0 \\\hline
		0 & 0 & 0 & 0 & 0 & 0 & 1 & 1 & 0 & 1 & 1 & 0 \\
		0 & 0 & 0 & 0 & 0 & 0 & 1 & 0 & 0 & 1 & 0 & 0 \\
		0 & 0 & 0 & 0 & 0 & 0 & 1 & 0 & 1 & 1 & 0 & 1 \\\hline
		0 & 0 & 0 & 0 & 0 & 0 & 0 & 0 & 0 & 1 & 1 & 0 \\
		0 & 0 & 0 & 0 & 0 & 0 & 0 & 0 & 0 & 1 & 0 & 0 \\
		0 & 0 & 0 & 0 & 0 & 0 & 0 & 0 & 0 & 1 & 0 & 1 \\
	\end{array}\right)
	\quad\quad
	\left(\begin{array}{ccc|ccc|ccc|ccc}
		1 & 0 & 1 & 1 & 0 & 1 & 0 & 1 & 0 & 1 & 0 & 1 \\
		0 & 1 & 1 & 0 & 1 & 1 & 1 & 0 & 1 & 0 & 1 & 1 \\
		1 & 0 & 0 & 1 & 0 & 0 & 1 & 1 & 0 & 1 & 0 & 0 \\\hline
		0 & 0 & 0 & 1 & 0 & 1 & 1 & 0 & 1 & 0 & 1 & 0 \\
		0 & 0 & 0 & 0 & 1 & 1 & 0 & 1 & 1 & 1 & 0 & 1 \\
		0 & 0 & 0 & 1 & 0 & 0 & 1 & 0 & 0 & 1 & 1 & 0 \\\hline
		0 & 0 & 0 & 0 & 0 & 0 & 1 & 0 & 1 & 1 & 0 & 1 \\
		0 & 0 & 0 & 0 & 0 & 0 & 0 & 1 & 1 & 0 & 1 & 1 \\
		0 & 0 & 0 & 0 & 0 & 0 & 1 & 0 & 0 & 1 & 0 & 0 \\\hline
		0 & 0 & 0 & 0 & 0 & 0 & 0 & 0 & 0 & 1 & 0 & 1 \\
		0 & 0 & 0 & 0 & 0 & 0 & 0 & 0 & 0 & 0 & 1 & 1 \\
		0 & 0 & 0 & 0 & 0 & 0 & 0 & 0 & 0 & 1 & 0 & 0 \\
	\end{array}\right)
    }
	$}
\]
\vspace{1em}
\noindent Matrices \(M_7, M_8, M_9\):
\[
\scalebox{0.95}{$
	{\renewcommand{\arraystretch}{0.46}\setlength{\arraycolsep}{1.5pt}
	\left(\begin{array}{ccc|ccc|ccc|ccc}
		0 & 1 & 0 & 0 & 0 & 0 & 1 & 0 & 0 & 1 & 1 & 0 \\
		1 & 1 & 0 & 0 & 0 & 0 & 0 & 1 & 1 & 1 & 0 & 1 \\
		0 & 1 & 1 & 0 & 0 & 0 & 0 & 1 & 0 & 0 & 0 & 1 \\\hline
		0 & 0 & 0 & 0 & 1 & 0 & 0 & 0 & 0 & 1 & 0 & 0 \\
		0 & 0 & 0 & 1 & 1 & 0 & 0 & 0 & 0 & 0 & 1 & 1 \\
		0 & 0 & 0 & 0 & 1 & 1 & 0 & 0 & 0 & 0 & 1 & 0 \\\hline
		0 & 0 & 0 & 0 & 0 & 0 & 0 & 1 & 0 & 0 & 0 & 0 \\
		0 & 0 & 0 & 0 & 0 & 0 & 1 & 1 & 0 & 0 & 0 & 0 \\
		0 & 0 & 0 & 0 & 0 & 0 & 0 & 1 & 1 & 0 & 0 & 0 \\\hline
		0 & 0 & 0 & 0 & 0 & 0 & 0 & 0 & 0 & 0 & 1 & 0 \\
		0 & 0 & 0 & 0 & 0 & 0 & 0 & 0 & 0 & 1 & 1 & 0 \\
		0 & 0 & 0 & 0 & 0 & 0 & 0 & 0 & 0 & 0 & 1 & 1 \\
	\end{array}\right)
	\quad\quad
	\left(\begin{array}{ccc|ccc|ccc|ccc}
		0 & 1 & 1 & 0 & 0 & 0 & 0 & 1 & 0 & 0 & 0 & 1 \\
		0 & 0 & 1 & 0 & 0 & 0 & 1 & 1 & 0 & 1 & 1 & 1 \\
		1 & 1 & 0 & 0 & 0 & 0 & 0 & 1 & 1 & 1 & 0 & 1 \\\hline
		0 & 0 & 0 & 0 & 1 & 1 & 0 & 0 & 0 & 0 & 1 & 0 \\
		0 & 0 & 0 & 0 & 0 & 1 & 0 & 0 & 0 & 1 & 1 & 0 \\
		0 & 0 & 0 & 1 & 1 & 0 & 0 & 0 & 0 & 0 & 1 & 1 \\\hline
		0 & 0 & 0 & 0 & 0 & 0 & 0 & 1 & 1 & 0 & 0 & 0 \\
		0 & 0 & 0 & 0 & 0 & 0 & 0 & 0 & 1 & 0 & 0 & 0 \\
		0 & 0 & 0 & 0 & 0 & 0 & 1 & 1 & 0 & 0 & 0 & 0 \\\hline
		0 & 0 & 0 & 0 & 0 & 0 & 0 & 0 & 0 & 0 & 1 & 1 \\
		0 & 0 & 0 & 0 & 0 & 0 & 0 & 0 & 0 & 0 & 0 & 1 \\
		0 & 0 & 0 & 0 & 0 & 0 & 0 & 0 & 0 & 1 & 1 & 0 \\
	\end{array}\right)
	\quad\quad
	\left(\begin{array}{ccc|ccc|ccc|ccc}
		1 & 1 & 0 & 0 & 0 & 0 & 0 & 1 & 1 & 1 & 0 & 1 \\
		1 & 0 & 1 & 0 & 0 & 0 & 0 & 0 & 1 & 1 & 0 & 0 \\
		0 & 0 & 1 & 0 & 0 & 0 & 1 & 1 & 0 & 1 & 1 & 1 \\\hline
		0 & 0 & 0 & 1 & 1 & 0 & 0 & 0 & 0 & 0 & 1 & 1 \\
		0 & 0 & 0 & 1 & 0 & 1 & 0 & 0 & 0 & 0 & 0 & 1 \\
		0 & 0 & 0 & 0 & 0 & 1 & 0 & 0 & 0 & 1 & 1 & 0 \\\hline
		0 & 0 & 0 & 0 & 0 & 0 & 1 & 1 & 0 & 0 & 0 & 0 \\
		0 & 0 & 0 & 0 & 0 & 0 & 1 & 0 & 1 & 0 & 0 & 0 \\
		0 & 0 & 0 & 0 & 0 & 0 & 0 & 0 & 1 & 0 & 0 & 0 \\\hline
		0 & 0 & 0 & 0 & 0 & 0 & 0 & 0 & 0 & 1 & 1 & 0 \\
		0 & 0 & 0 & 0 & 0 & 0 & 0 & 0 & 0 & 1 & 0 & 1 \\
		0 & 0 & 0 & 0 & 0 & 0 & 0 & 0 & 0 & 0 & 0 & 1 \\
	\end{array}\right)
    }
	$}
\]


\begingroup

\begin{thebibliography}{99}

\bibitem{silva2008}
Silva, D., Kschischang, F.R., K{\"o}tter, R.: A rank-metric approach to error control in random network coding. IEEE Trans. Inf. Theory \textbf{54}(9), 3951--3967 (2008).

\bibitem{delsarte}
Delsarte, P.: Bilinear forms over a finite field, with applications to coding theory. J. Combin. Theory Ser. A \textbf{25}(3), 226--241 (1978).

\bibitem{gabidulin}
Gabidulin, E.M.: Theory of codes with maximum rank distance. Probl. Inf. Transm. \textbf{21}(1), 1--12 (1985).

\bibitem{Neri2019}
Neri, A.: Algebraic Theory of Rank-Metric Codes: Representations, Invariants and Density Results. Ph.D. thesis, University of Zurich (2019).

\bibitem{neri2024}
Neri, A., Stanojkovski, M.: A proof of the Etzion--Silberstein conjecture for monotone and MDS-constructible Ferrers diagrams. J. Combin. Theory Ser. A \textbf{208}, 105937 (2024).

\bibitem{EtzionSilberstein2013}
Etzion, T., Silberstein, N.: Codes and designs related to lifted MRD codes. IEEE Trans. Inf. Theory \textbf{59}(2), 1004--1017 (2013).

\bibitem{etzion2009}
Etzion, T., Silberstein, N.: Error-correcting codes in projective spaces via rank-metric codes and Ferrers diagrams. IEEE Trans. Inf. Theory \textbf{55}(7), 2909--2919 (2009).

\bibitem{etzion2016}
Etzion, T., Gorla, E., Ravagnani, A., Wachter-Zeh, A.: Optimal Ferrers diagram rank-metric codes. IEEE Trans. Inf. Theory \textbf{62}(4), 1616--1630 (2016).

\bibitem{liu2019}
Liu, S., Chang, Y., Feng, T.: Constructions for optimal Ferrers diagram rank-metric codes. IEEE Trans. Inf. Theory \textbf{65}(7), 4115--4130 (2019).

\bibitem{liu2019several}
Liu, S., Chang, Y., Feng, T.: Several classes of optimal Ferrers diagram rank-metric codes. Linear Algebra Appl. \textbf{581}, 128--144 (2019).

\bibitem{liu2023}
Liu, S.: Optimal Ferrers diagram rank-metric codes from MRD codes. Des. Codes Cryptogr. \textbf{91}(12), 3977--3993 (2023).

\bibitem{liu2020parallel}
Liu, S., Chang, Y., Feng, T.: Parallel multilevel constructions for constant dimension codes. IEEE Trans. Inf. Theory \textbf{66}(11), 6884--6897 (2020).

\bibitem{liu2023double}
Liu, S., Ji, L.: Double multilevel constructions for constant dimension codes. IEEE Trans. Inf. Theory \textbf{69}(1), 157--168 (2023).

\bibitem{antrobus2019}
Antrobus, J., Gluesing-Luerssen, H.: Maximal Ferrers diagram codes: Constructions and genericity considerations. IEEE Trans. Inf. Theory \textbf{65}(10), 6204--6223 (2019).

\bibitem{pratihar2023}
Pratihar, R., Randrianarisoa, T.H.: Constructions of optimal rank-metric codes from automorphisms of rational function fields. Adv. Math. Commun. \textbf{17}(1), 262--287 (2023).

\bibitem{calderini2026}
Calderini, M., Messia, M., Neri, A.: On the Etzion--Silberstein conjecture for block Ferrers diagrams. arXiv:2607.08239v1 (2026).

\bibitem{zhang2019}
Zhang, T., Ge, G.: Constructions of optimal Ferrers diagram rank metric codes. Des. Codes Cryptogr. \textbf{87}(1), 107--121 (2019).

\bibitem{fu2026}
Fu, F.-W., Gao, X., Mesnager, S., Wang, G.: Optimal Ferrers diagram rank-metric codes: New constructions, diagram combinations, and applications to constant-dimension subspace codes. arXiv:2609.24239v1 (2026).

\bibitem{koetter2008}
K{\"o}tter, R., Kschischang, F.R.: Coding for errors and erasures in random network coding. IEEE Trans. Inf. Theory \textbf{54}(8), 3579--3591 (2008).

\bibitem{couvee2026}
Beeloo-Sauerbier Couv\'ee, H., Neri, A.: Irreducible Ferrers diagrams in the Etzion--Silberstein conjecture. arXiv:2604.27868v1 (2026).

\bibitem{xu2025}
Xu, D., Li, M.: Multilevel constructions of constant dimension codes based on one-factorization of complete graphs. Cryptogr. Commun. \textbf{18}, 2409--2428 (2026).

\bibitem{XuSong2025}
Xu, D., Song, Y.: Coset constructions of constant dimension codes by cosets of optimal Ferrers diagrams rank metric codes. IEEE Trans. Inf. Theory \textbf{71}(10), 7959--7975 (2025).

\bibitem{hong2025}
Hong, X., Cao, X., Luo, G.: Generalized bilateral multilevel construction for constant dimension codes. Des. Codes Cryptogr. \textbf{93}(1), 197--225 (2025).

\bibitem{silberstein2015}
Silberstein, N., Trautmann, A.-L.: Subspace codes based on graph matchings, Ferrers diagrams, and pending blocks. IEEE Trans. Inf. Theory \textbf{61}(7), 3937--3953 (2015).

\bibitem{yu2022bilateral}
Yu, S., Ji, L., Liu, S.: Bilateral multilevel construction of constant dimension codes. Adv. Math. Commun. \textbf{16}(4), 1165--1183 (2022).

\bibitem{GowQuinlan2009}
Gow, R., Quinlan, R.: Galois extensions and subspaces of alternating bilinear forms with special rank properties.
Linear Algebra Appl. \textbf{430}(8-9), 2212--2224 (2009).

\bibitem{Prajapati2026}
Prajapati, J.: A counterexample to the Etzion-Silberstein conjecture.
arXiv:2608.08478v1 (2026). 


\end{thebibliography}
\endgroup
\end{document}